\documentclass[useAMS,useusegraphicx]{mn2e}
\usepackage{amssymb}
\usepackage{aas_macros}
\usepackage{epic,overpic}
\usepackage{url}
\usepackage[usenames,dvipsnames]{color}
\definecolor{lightyellow}{rgb}{1,1,0.7}
\newcommand{\be}{\begin{equation}}
\newcommand{\ee}{\end{equation}}
\newcommand{\ba}{\begin{eqnarray}}
\newcommand{\ea}{\end{eqnarray}}

\def\dlsds{d_{\rm ls}/d_{\rm s}}

\title[WL Detection of the Distance-Redshift
Relation]{A Weak Lensing Detection of the Cosmological
  Distance-Redshift Relation Behind Three Massive
  Clusters\thanks{Based on data collected at Subaru Telescope and
    obtained from the SMOKA, which is operated by the Astronomy Data
    Center, National Astronomical Observatory of Japan.}}

\author[Medezinski et al.
2010]{Elinor Medezinski$^{1}$, Tom
  Broadhurst$^{2,3}$, Keiichi Umetsu$^{4}$, Narciso Ben\'\i tez$^{5}$
\newauthor \& Andy Taylor$^{6}$\\
  $^{1}${Department of Physics \& Astronomy, The Johns Hopkins University, 3400
N. Charles Street, Baltimore, MD 21218, USA} \\
  $^{2}${Theoretical physics, University of the Basque Country,
       Bilbao 48080, Spain}\\
  $^{3}${Ikerbasque, Basque Foundation for Science, Alameda Urquijo, 36-5
Plaza Bizkaia 48011, Bilbao, Spain}\\
  $^{4}${Institute of Astronomy and Astrophysics, Academia Sinica,
    P.~O.  Box 23-141, Taipei 106, Taiwan, Republic of
    China} \\
  $^{5}${Instituto de Astrof\'\i sica de Andaluc\'\i a (CSIC),
    Granada, Spain} \\
  $^{6}${Scottish Universities Physics Alliance (SUPA), Institute for
    Astronomy, School of Physics, University of Edinburgh,}\\
    { Royal Observatory, Blackford Hill, Edinburgh, EH9 3HJ, U.K.}  }

\begin{document}
\maketitle

\begin{abstract}
  The amplitude of weak lensing should increase with source distance,
rising steeply behind a lens and saturating at high redshift,
providing a model-independent means of measuring cosmic geometry. We
measure the amplitude of weak lensing with redshift for three massive
clusters, A370 ($z=0.375$), ZwCl0024+17 ($z=0.395$) and RXJ1347-11
($z=0.451$), using deep, three-colour Subaru imaging. We define the depth of
lensed populations with reference to the COSMOS and GOODS fields,
providing a consistency check of photo-z estimates over a wide range
of redshift and magnitude. The predicted distance-redshift relation is
followed well for the deepest dataset, A370, for a wide range of
cosmologies, and is consistent with less accurate data for the other
two clusters.  Scaling this result to a new survey of $\sim25$ massive
clusters should provide a useful cosmological constraint on $w$,
complementing existing techniques, with distance measurements covering
the untested redshift range, $1<z<5$.
\end{abstract}



\section{Introduction}\label{Intro}

Constraining cosmological parameters has been the focus of major
surveys in the last decade, via precision cosmic microwave background
(CMB) temperature correlations
\nocite{2007ApJS..170..377S,2009ApJ...705..978B}({Spergel} {et~al.} 2007; {Brown} {et~al.} 2009) and SN-Ia light curves
\nocite{1998AJ....116.1009R,1999ApJ...517..565P}({Riess} {et~al.} 1998; {Perlmutter} {et~al.} 1999). A standard,
$\Lambda$CDM, cosmological model has been defined by this work, albeit at the
price of accepting an accelerating expansion driven by a cosmological
constant, and non-baryonic dark matter (DM) of an unknown nature
dominating the mass density of the Universe.  Measurements of the
angular diameter distance of the CMB refer to $z\sim1100$, and
luminosity distances are derived from SN-Ia in the range $z<1$. In
principle, lensing can provide a complementary distance measurements
in the range, $z>1$, from the purely geometric deflection of light,
which increases with source distance behind a lens.

For lensing clusters, the bend-angle of light scales linearly with
angular diameter distance ratio, $\dlsds$, the separation between the
lens and the source, divided by distance to the source. This distance
ratio has a characteristic geometric dependence on redshift, rising
steeply behind the lens and then saturating at large source redshift
\nocite{1995ApJ...438...49B}(e.g., Fig.~1 of {Broadhurst}, {Taylor} \&  {Peacock} 1995).  This effect has been
detected behind massive lensing clusters, where the separation in
angle between multiple images of higher redshift sources is noticeably
larger than for lower redshift sources.  For example, for the well
studied cluster SDSSJ1004+4112 ($z=0.68$), five images of a QSO at
$z=1.734$
are within
an Einstein radius of about $\theta_E\sim7\arcsec$
\nocite{2003Natur.426..810I,2004ApJ...605...78O}({Inada} {et~al.} 2003; {Oguri} {et~al.} 2004), whereas a more
distant multiply lensed galaxy behind this cluster at
$z=3.332$ is at
a much larger Einstein radius of $\theta_E\sim16\arcsec$
\nocite{2005ApJ...629L..73S}({Sharon} {et~al.} 2005). Other sets of multiple images show this
increasing angular scaling with source redshift, following well the
expected general form of the redshift-distance relation in careful
strong lensing analyses of deep Hubble data
\nocite{2004A&A...417L..33S,2005ApJ...621...53B,2009MNRAS.396.1985Z,
2009ApJ...703L.132Z,2009ApJ...707L.102Z}({Soucail}, {Kneib} \&  {Golse} 2004; {Broadhurst} {et~al.} 2005; {Zitrin} {et~al.} 2009b; {Zitrin} \& {Broadhurst} 2009; {Zitrin} {et~al.} 2009a).

However, these studies are not able to distinguish between the
relatively subtle changes between cosmologies in the range of
interest, due to the inherent insensitivity of the distance ratio
$\dlsds$ to the cosmological parameters. Moreover, the bend-angle is
particularly sensitive to the gradient of the mass profile, requiring
many sets of multiple images in the strong regime to simultaneously
solve for both the cosmological model and the mass distribution. Instead in
practice, strong lens modeling usually adopts the standard
cosmological relation in order to better derive the mass distribution,
with multiply lensed sources forced to lie on the lensing
distance-redshift relation. This helps eliminate the otherwise
considerable degeneracy in constraining the slope of the lensing mass
profile of a galaxy cluster
\nocite{2005ApJ...621...53B,2009MNRAS.396.1985Z,2009ApJ...703L.132Z,
2009ApJ...707L.102Z}({Broadhurst} {et~al.} 2005; {Zitrin} {et~al.} 2009b; {Zitrin} \& {Broadhurst} 2009; {Zitrin} {et~al.} 2009a).
The hope of constraining the cosmography from strong lensing data is
remote with current tools \nocite{2009MNRAS.396..354G}({Gilmore} \& {Natarajan} 2009).

Weak lensing (WL), by contrast, offers a model independent way of constraining
the cosmological parameters via the distance-redshift relation
\nocite{2004MNRAS.353.1176T}({Taylor} {et~al.} 2004). Image distortions and magnification depend on
gradients of the deflection field and in the WL limit, these are just
proportional to $\dlsds$. The mass profile enters only in the stronger regime as
a second order correction \nocite{2007ApJ...663..717M}({Medezinski} {et~al.} 2007).  Here we are concerned
with the amplitude of cluster WL dependence on source redshift, for which no
mass reconstruction is required when evaluating the distance-redshift relation.
However, although this observed effect is independent of the mass distribution
in the weak limit, the sensitivity to cosmological parameters is still
inherently very small.

The analogous effect in the field has been explored in terms of the cosmic
shear, to measure the general mass distribution. Optimal formalisms have been
developed which cross-correlate the foreground distribution of galaxies along a
given line of sight with the distribution of background images
\nocite{2001ApJ...557L..89W,2003PhRvL..91n1302J,2003MNRAS.344..673B,
2004MNRAS.353.1176T,2007MNRAS.374.1377T}({Wittman} {et~al.} 2001; {Jain} \& {Taylor} 2003; {Bacon} {et~al.} 2003; {Taylor} {et~al.} 2004, 2007) with clear detections of large scale
structure, including the COMBO-17 fields
\nocite{2003MNRAS.341..100B,2007MNRAS.376..771K}({Brown} {et~al.} 2003; {Kitching} {et~al.} 2007) and the COSMOS field
\nocite{2007ApJS..172..239M,2010A&A...516A..63S}({Massey} {et~al.} 2007; {Schrabback} {et~al.} 2010). To usefully derive cosmological parameters from
general cosmic shear work, deep all-sky surveys have been proposed (e.g.,
LSST\footnote{\url{http://www.lsst.org/lsst}},
DES\footnote{\url{https://www.darkenergysurvey.org/}},
JDEM\footnote{\url{http://jdem.gsfc.nasa.gov}}, EUCLID\footnote{\url{
http://sci.esa.int/science-e/www/object/index.cfm?fobjectid=42266}}).

Here we make use of detailed colour-colour (CC) information for three
intermediate redshift lensing clusters, A370 ($z=0.375$), ZwCl0024+17
($z=0.395$), RXJ1347-11 ($z=0.451$) and define several samples of galaxies of
differing background depths with which to explore the dependence of WL
distortion on source redshift.
With only three bands we cannot
  reliably define photometric redshifts for a sizable proportion of
  objects, but instead, by reference to the redshift
  disributions of the well studied deep field surveys, we may reliably
  define several galaxy populations of differing mean depths in
  the CC-space covered by the filters used for each cluster.

  Three colour selection has also been applied in a similar context to
  simulations aimed at forecasting the capabilities of WL
  tomograghy \nocite{2007JCAP...03..013J,2010MNRAS.405..257M}({Jain}, {Connolly} \&  {Takada} 2007; {Medezinski} {et~al.} 2010).  These
  simulations convincingly demonstrate that greater efficiency is
  likely by using limited 3-band imaging for WL tomography, rather than
  by investing greater imaging time in additional bands to improve
  photometric redshift precision.  For our purposes too we show here
  that a judicious choice of non-orthogonal boundaries in CC-space allows
  the definition of several distinct redshift samples of
  differing mean depth, with relatively little overlap in redshift.

Here we rely on well studied deep field surveys to estimate
the redshift distribution of these different background populations,
using the wide-field COSMOS 30-band photometric redshift survey
\nocite{2009ApJ...690.1236I}({Ilbert} {et~al.} 2009) and also the deeper GOODS-MUSIC survey
which has wide multi-wavelength coverage in 14 bands (from the U band to the Spitzer $8~\mu$m band)  \nocite{2006A&A...449..951G}({Grazian} {et~al.} 2006).

In \S~\ref{sec:data} we present the cluster observations and data
reduction and in \S~\ref{sec:samples} we explain the selection of
background galaxy samples and in \S~\ref{sec:wl} we describe the WL
analysis and outline the formalism. In \S~\ref{sec:results} we
derive the WL amplitude and mean redshift information of the
background samples, presenting our results regarding the lensing
distance-redshift relation. We discuss the requirements for
constraining the cosmological model with this method in \ref{sec:cosmo}.  We
summarize and conclude in \S~\ref{sec:summary}.

\section{Subaru Data reduction}  \label{sec:data}

\begin{table}
  \caption{The Cluster Sample: Redshift and Filter Information}
  \begin{tabular}{@{\hspace{0.01cm}}cccc@{\hspace{0.01cm}}}
    \hline
    {Cluster} & {$z$} & {Filters used$^1$} & {Seeing}\\
    & &(exp' time in sec) & (arcsec) \\
    \hline
    A370 & 0.375 & $B_{\rm J}(7200),{\bf R_{\rm C}}(8340),z'(14221)$ & 0.6\\
    ZwCl0024+17 & 0.395 & $B_{\rm J}(3600),{\mathbf R_{\rm C}}(5280),z'(1680)$ &
0.8\\
    RXJ1347-11 & 0.451 & $V_{\rm J}(1800),{\bf R_{\rm C}}(2880),z'(4860)$ &
0.76\\
    \hline\label{tab:cluster}
  \end{tabular}
\flushleft\footnotesize{$^1$Detection band marked in bold.}
\end{table}

We analyze deep images of three intermediate-redshift clusters, A370
ZwCl0024+17 and RXJ1347-11, observed with the wide-field camera
Suprime-Cam \nocite{2002PASJ...54..833M}({Miyazaki} {et~al.} 2002) in several optical bands, at
the prime focus of the 8.3m Subaru telescope.  The clusters are
publicly available from the Subaru archive,
SMOKA\footnote{\url{http://smoka.nao.ac.jp}}. Subaru reduction software
(SDFRED) developed by \nocite{2002AJ....123...66Y}{Yagi} {et~al.} (2002) is used for
flat-fielding, instrumental distortion correction, differential
refraction, sky subtraction and stacking.  Photometric catalogs are
created using {\sc SExtractor} \nocite{1996A&AS..117..393B}({Bertin} \& {Arnouts} 1996). Since our
work relies much on the colours of galaxies, we prefer using isophotal
magnitudes. We use the {\sc Colorpro} \nocite{2006AJ....132..926C}({Coe} {et~al.} 2006)
program to detect in the $R_{\rm C}$-band and measure colours through
matched isophotes in the other two bands. Astrometric correction is
done with {\sc Scamp} \nocite{2006ASPC..351..112B}({Bertin} 2006) using reference
objects in the {\sc NOMAD} catalogue \nocite{2004AAS...205.4815Z}({Zacharias} {et~al.} 2004) and the
SDSS-DR6 \nocite{2008ApJS..175..297A}({Adelman-McCarthy} {et~al.} 2008) where available. The
observational details are listed in Table~\ref{tab:cluster}.

\section{Sample Selection from the colour-colour diagram}
\label{sec:samples}

For each cluster we use Subaru observations in three broad optical
passbands and all observations are of good seeing, representing some
of the highest quality imaging by Subaru in terms of depth,
resolution, and colour coverage.  We first describe how we separate
the background galaxies from foreground and cluster galaxies,
combining WL measurements and the distribution of objects in the CC
plane and their clustering relative to the center of the cluster. We
then examine the redshift distribution of objects selected to lie in
the background using the CC plane with reference to the COSMOS field
where deep photometric redshifts are established to faint limits using
30 independent passbands covering a very wide range of wavelength.

Note that with only three bands, only a small fraction of the objects have well
defined photometric redshifts, in the sense of having a single, narrow peak in
their probability $p(z)$, and therefore their use to separate among different
redshift populations is very limited. Using the $BRz'$ CC space in this way with
reference to the now well established redshift surveys is arguably more reliable
in separating galaxy populations of differing depths, in agreement with the
simulations of \nocite{2007JCAP...03..013J}{Jain} {et~al.} (2007).

\begin{figure*}
  \centering
  \includegraphics[width=8cm,height=6cm]{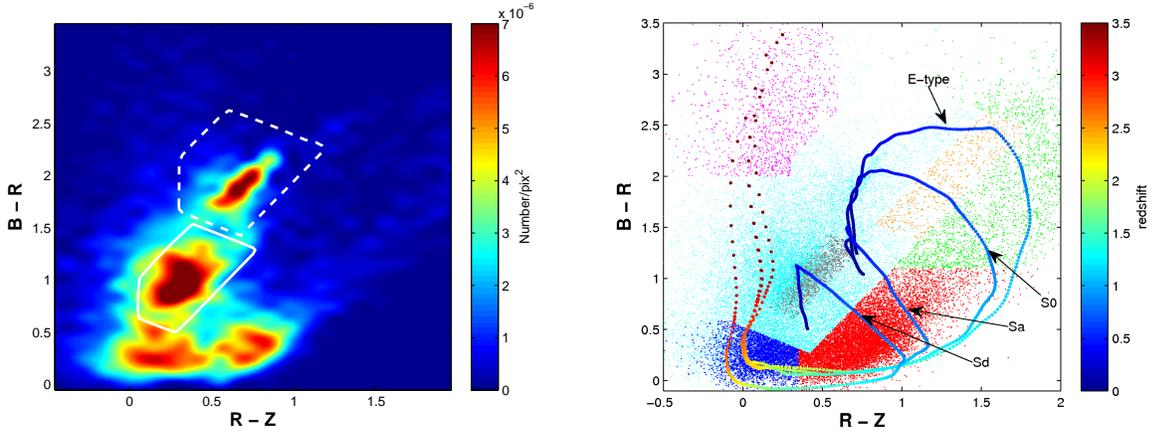}
 \includegraphics[width=8cm,height=6cm]{A370/A370_colorcolor_bg_evtracks_100820.eps3}
  \caption{{\it Left:} Number density in $B_{\rm J}-R_{\rm C}$ vs. $R_{\rm
C}-z'$ CC space for A370. The four distinct density peaks are shown to be
different galaxy populations - the reddest peak in the upper right corner of the
plots (dashed white line) depicts the overdensity of cluster galaxies, whose
colours lay on the red sequence; the middle peak lying blueward of the cluster
comprises mainly foreground galaxies (solid white line); the two peaks in the
bottom part (bluest in $B_{\rm J}-R_{\rm C}$) can be demonstrated to comprise of
blue and red background galaxies.  Together, the red+blue galaxies will serve as
our ``reference'' background sample for WL purposes. {\it Right:} $B_{\rm
J}-R_{\rm C}$ vs. $R_{\rm C}-z'$ CC diagram, showing the distribution of
galaxies in A370. Marked are the selected foreground sample (gray) and
background samples: orange (orange), green (green), red (red), blue (blue) and
dropout (magenta) background galaxies, selected to include galaxies lying away
from the cluster and foreground regions. Overlaid are synthetic colour tracks
including evolution, calculated with the Galev code for an elliptical, S0, Sa
and Sd type models.}
  \label{fig:CC}
\end{figure*}

In our previous analysis of these data, \nocite{2010MNRAS.405..257M}{Medezinski} {et~al.} (2010, hereafter
M10), we demonstrated how the cluster and
foreground galaxies can be reliably identified and separated from
background galaxies in the CC diagram, using $B_{\rm J},R_{\rm C},z'$
bands (Fig.~\ref{fig:CC}).  We found in the field of A370 the
prominent overdensity of galaxies centered on $B_{\rm J}-R_{\rm
C}\sim2$ and $R_{\rm C}-z'\sim0.8$ denotes the red-sequence of cluster
galaxies (see Fig.~\ref{fig:CC}, left-hand panel, where this
overdensity is enclosed by dashed white line). This was also shown by
the relatively small mean distance from cluster center of galaxies in
that region in CC space (Fig.~1 in M10). The WL measurements for this
population of objects is very close to zero, with a measured
tangential distortion profile, $g_T(r)$, consistent with zero all the
way out to the virial radius, as expected for cluster galaxies which
are unlensed.
The main central overdensity around $B_{\rm J}-R_{\rm C}\sim1$ and
$R_{\rm C}-z'\sim0.3$ consists of many foreground galaxies (see
Fig.~\ref{fig:CC}, left-hand panel, with an overdensity enclosed by
solid white line). These galaxies show a very low level $g_T$ compared
to the reference background, marked in gray on Fig.~\ref{fig:CC}
(right panel), and their surface density profile shows only modest central
clustering (see M10) indicating that most of these objects
lie in the foreground of the cluster.
When selecting background galaxies we stay well away from these regions
of CC space which are dominated by cluster and foreground galaxies
to minimize contamination by these unlensed galaxies, as described below.
%

To identify background populations in M10, we took into account both
the WL signal and the density distribution of galaxies in the CC
plane.  In CC space a relatively red population can be rather well
defined, dominated by an obvious overdensity (around $B_{\rm J}-R_{\rm
  C}\sim0.5$ and $R_{\rm C}-z'\sim0.8$), and the bluest population is
confined to a separate cloud (around $B_{\rm J}-R_{\rm C}\sim0.3$ and
$R_{\rm C}-z'\sim0.2$) of faint galaxies with a clear lensing signal.
Looking at the WL profiles of the red and blue samples, we see very
similar behaviour, with a continuous rising signal toward the cluster
center. Both cases show good agreement with each other (M10).
Combined together, they form our \textbf{reference background} sample,
to which all the other samples we derive below will be normalized.


The validity of our selection was also demonstrated in M10 by
comparison with the spectral evolution of galaxies calculated with the
stellar synthesis code Galev\footnote{\url{http://www.galev.org/}}
\nocite{2009MNRAS.396..462K}({Kotulla} {et~al.} 2009). Here we overlay the CC diagram with
colour-tracks of galaxy models: E-type (exponentially declining SFR
with $Z_\odot$), S0 (gas-related SFR with $Z_\odot$), Sa (gas-related
SFR with $2.5Z_\odot$), and Sd (constant SFR with $0.2Z_\odot$).
These evolutionary tracks originate in the low-redshift cloud we
established as foreground in CC space and evolve to higher redshift
passing through the cluster redshift and then to bluer colours as shown
in Fig.~\ref{fig:z_CC} (right), and finally end at the top-left corner
of our CC space, dropping out of the $B_{\rm J}$-band at a redshift of
$z\sim3.5$

In
this paper we add additional samples of background galaxies.
Two samples will consist of galaxies at redshifts not far beyond that
of the cluster, which we term ``orange'' and ``green''. These are selected to
lie on
the upper-right of the CC space, corresponding reasonably to colour-tracks of
E/S0 galaxies which are predicted to show a bend in the CC plane,
becoming bluer ($B_{\rm J}-R_{\rm C}\sim2-2.5$ and $R_{\rm
  C}-z'\sim1-1.5$) toward higher redshift, matching well our observed
distribution.  A third group comprises the red-cloud described above,
centered on $B_{\rm J}-R_{\rm C}\sim0.5$ and $R_{\rm C}-z'\sim0.8$, we
call the ``red'' sample. This sample consists of an overdensity of
background galaxies from the known ``red'' branch of the bimodality of
field galaxies colours \nocite{2007ApJS..172...99C}({Capak} {et~al.} 2007). A Forth sample
consists of the prominent blue peak identified as background galaxies,
lying at redshifts beyond the red galaxies, corresponding to the
``blue'' branch of the colour bimodality, where the colour-tracks turn
redder at $B_{\rm J}-R_{\rm C}$ as the UV-flux starts to drop out of
the blue $B_{\rm J}$ band with increasing redshift, above a redshift
of approximately, $z\sim2$.  Finally, we select galaxies from the
top-left corner of the CC plane, corresponding to the predicted
location of $B_{\rm J}$-dropout galaxies at an average redshift of
$z\sim 3.5$.

For each sample selected we examine the colour boundaries of
each sample as a function of the WL signal, to ensure we keep well
clear of foreground or cluster members which otherwise dilute the WL
signal of background galaxies. This approach has been investigated in
our earlier work, where this problem was first identified as a major
problem for WL work and rectified using appropriate CC-selection
\nocite{2007ApJ...663..717M,2008ApJ...684..177U,2010MNRAS.405..257M,
2010ApJ...714.1470U}({Medezinski} {et~al.} 2007; {Umetsu} \& {Broadhurst} 2008; {Medezinski} {et~al.} 2010; {Umetsu} {et~al.} 2010).

We apply this selection scheme for two more clusters, ZwCl0024+17 and
RXJ1347-11, though here the data is not as deep. Furthermore, the colour
coverage is less wide in the case of RXJ1347-11 (only $VRz'$) allowing the
selection of the corresponding orange, green, red and blue background samples,
as above, but not sufficient for  separation of dropout galaxies for a
significant WL detection.

The greater depth of the A370 imaging allows us to further divide the
green, red and blue background samples into independent bright and faint
subsamples for our WL measurements.
For each of these samples, we now estimate their average WL signal, by
comparing the tangential distortion measurements to that of the
reference background sample, and derive what we call mean ``$g_T$
amplitude'' (see \S~\ref{sec:wl-prof}). Subsequently, we also estimate
the median redshift from the COSMOS photo-$z$ catalogue and its equivalent
$\dlsds$ (see \S~\ref{sec:cosmos-z}). We summarize the properties of
the selected samples in Table \ref{tab:samples}.

\begin{table*}
  \centering
  \caption{CC-selected Sample Properties}
  \begin{tabular}{cccccccc}
    \hline
    {Cluster} & {Sample} & {magnitude limits} &  {N} &{$\bar{n}$} &
    {$\Gamma$ } &  $\chi^2/dof$ & {$<{z_s}>$}\\
       & & & & arcmin$^{-2}$ &{\tiny ($g_T$-amplitude ratio)} & {\tiny (PL)}&\\
     [0.5ex]\hline\\[-1.8ex]
     A370 & foreground  & $18< z'<22$    & 1235& 1.3&-0.04&13/6 &0.33 \\
	  & orange      & $19< z'<23$    & 450 & 0.5&0.41&  1/6&0.73 \\
	  & green-bright& $20< z'<22.5$  & 839 & 0.9&0.85& 11/6&0.9 \\
          & green-faint & $22.5< z'<25$  & 1240& 1.3&0.88&  6/7&0.99\\
          & red-bright  & $22< z'< 24$   & 5454& 5.6&0.87& 17/8&1.11 \\
          & red-faint   & $24< z'< 26$   & 6986& 7.1&1.1 & 27/8&1.14\\
          & blue-bright & $23< z'< 24.6$ & 1679& 1.7&0.95&  6/8&1.79\\
          & blue-faint  &$24.6< z'< 25.5$& 2857& 2.9&1.15& 23/8&1.78\\
          & drops       &$24< z'< 26.5$  & 1529& 1.6&1.3 & 13/8&3.86\\
          & background  & $22< z'< 26$   &19362&19.8&1   & 19/8&1.31\\
          \hline
ZwCl0024+17& foreground& $18< z'<22$   & 734  & 0.9&0.08&  5/9&0.29\\
	   & orange    & $18< z'<23$   & 687  & 0.9&0.55& 11/9&0.73\\
	   & green     & $21< z'<25$   & 1582 & 2  &0.83& 23/9&0.98\\
           & red       & $21< z'< 25.5$& 7716 & 9.7&0.97&  8/9&1.11\\
           & blue      & $23< z'< 25$  & 2420 & 3  &0.99& 13/9&1.62\\
           & drops     & $25< z'< 27$  &  625 & 0.6&0.93& 23/9&3.71\\
           & background& $21< z'< 25.5$&10488 &13.1&1   &  7/8&1.24\\
          \hline
RXJ1347-11& foreground& $19< z'<24$   & 2452 & 3.4 &-0.04&14/8&0.43\\
          & orange    & $19< z'<26$   &  473 & 0.5 & 0.76& 5/8&0.76\\
          & green     & $21< z'<25$   & 1296 & 1.1 & 1.1 & 5/8&0.94\\
          & red       & $21< z'< 26$  & 4880 & 3.9 & 0.92& 9/8&1.11\\
          & blue      & $23.5< z'< 26$& 1509 & 2.1 & 1.18&35/8&1.95\\
          & background& $21< z'< 26$  & 7139 & 6.3 & 1   & 4/6&1.24\\
          \noalign{\smallskip}\hline\label{tab:samples}
   \end{tabular}
\end{table*}

\section{Weak Lensing measurements} \label{sec:wl}

To make the WL catalogs, we use the IMCAT package developed by N.
Kaiser\footnote{\url{http://www.ifa.hawaii.edu/~kaiser/imcat}} to perform
object detection and shape measurements, following the formalism
outlined in \nocite{1995ApJ...449..460K}{Kaiser}, {Squires} \&  {Broadhurst} (1995, hereafter KSB).  Our analysis
pipeline is described in \nocite{2010ApJ...714.1470U}{Umetsu} {et~al.} (2010). We have tested
our shape measurement and object selection pipeline using STEP
\nocite{2006MNRAS.368.1323H}({Heymans} {et~al.} 2006) data of mock ground-based observations
(see \nocite{2010ApJ...714.1470U}{Umetsu} {et~al.} 2010, \S~3.2). Full details of the
methods are presented in \nocite{2008ApJ...684..177U}{Umetsu} \& {Broadhurst} (2008),
\nocite{2009ApJ...694.1643U}{Umetsu} {et~al.} (2009) and \nocite{2010ApJ...714.1470U}{Umetsu} {et~al.} (2010).

The shape distortion of an object is described by the complex
reduced-shear, $g=g_1+ig_2$, where the reduced-shear is defined as:
\begin{equation}
g_{\alpha}\equiv\gamma_{\alpha}/(1-\kappa).
\end{equation}
The tangential component $g_{T}$ is used to obtain the azimuthally
averaged distortion due to lensing, and computed from the distortion
coefficients $g_{1},g_{2}$:
\begin{equation}
g_{T}=-( g_{1}\cos2\theta +  g_{2}\sin2\theta),
\end{equation}
where $\theta$ is the position angle of an object with respect to the
cluster centre, and the uncertainty in the $g_{T}$ measurement is
$\sigma_{T} = \sigma_{g}/\sqrt{2}\equiv \sigma$ in terms of the RMS
error $\sigma_{g}$ for the complex shear measurement.  To improve the
statistical significance of the distortion measurement, we calculate
the weighted average of $g_{T}$ and its weighted error, as
\begin{eqnarray}\label{eq:mean_gt}
  \langle g_{T}(\theta_n)\rangle &=& \frac{\sum_i u_{g,i}\, g_{T,i}}
  {\sum_i u_{g,i}},\\
  \sigma_{T}(\theta_n) &=& \sqrt{
    \frac{\sum_i u^2_{g,i}\sigma^2_i}{\left(\sum_i u_{g,i}\right)^2},
  }
\end{eqnarray}
where the index $i$ runs over all of the objects located within the
$n$-th annulus with a median radius of $\theta_n$, and $u_{g,i}$ is
the inverse variance weight for $i$-th object,
$u_{g,i}=1/(\sigma_{g,i}^2+\alpha^2)$, where $\alpha^2$ is the
softening constant variance. We choose $\alpha=0.4$, which is a
typical value of the mean RMS $\bar{\sigma}_g$ over the background
sample.
 We accurately combine the photometry
  with weak-lensing measurements of as many galaxies as possible,
  discarding objects below the seeing limit (given in
  table~\ref{tab:cluster}) plus two standard deviation of that value
  in the detection band, to remove stars and avoid unreliable shape
  measurements.


\subsection{Formalism: Relative Distortion Strength}\label{sec:form}

For a given source redshift $z_s$ and a fixed lens redshift $z_l$, the
observable (complex) reduced gravitational shear $g(z_s)$ in the
subcritical regime is expressed in terms of the gravitational shear
$\gamma$ and the lens convergence $\kappa$ as
\nocite{1997A&A...318..687S,2007ApJ...663..717M}(e.g., {Seitz} \& {Schneider} 1997; {Medezinski} {et~al.} 2007)
\begin{equation}
g(z_s)=\gamma(z_s)(1-\kappa[z_s])^{-1}
=\gamma_{\infty}\,\sum_{k=0}^{\infty}
\beta^{k+1}(z_s)\kappa^k_{\infty}
\end{equation}
where $\kappa_{\infty}$ and $\gamma_{\infty}$ are the lensing
convergence and the gravitational shear, respectively, calculated for
a hypothetical source at $z_s\rightarrow\infty$, and $\beta(z_s)$ is the
lensing strength of a source at $z_s$ relative to a source at
$z_s\rightarrow\infty$, $\beta(z_s)\equiv D(z_s)/D(z_s\to \infty)$;
$D(z_s)\equiv\dlsds$.
Hence, the reduced shear averaged over the source redshift
distribution is expressed as
\begin{equation}\label{eq:mean_g}
  \langle g \rangle =
  \gamma_{\infty}\,\sum_{k=0}^{\infty} \langle \beta^{k+1}\rangle
  \kappa_{\infty}^k,
\end{equation}
where $\langle \beta^k \rangle$ is defined such that
\begin{equation}
  \langle \beta^k\rangle \equiv
  \frac{\int\!dz_s\,N(z_s)\beta^k(z_s)}{\int\!dz\,N(z_s)}
\end{equation}
with the redshift distribution $N(z_s)$.  In the WL limit
where $|\kappa_{\infty}|, |\gamma|_{\infty} \ll 1$, then
\begin{equation}\label{eq:gt-d}
  \langle g\rangle \approx \langle \beta \rangle
  \gamma_{\infty} = \langle\gamma\rangle.
\end{equation}
Thus, the mean reduced shear is simply proportional to the mean
lensing strength, $\langle \beta\rangle \propto \langle D\rangle$.  The
next order approximation is
\begin{equation}\label{eq:1storder}
  \langle g \rangle \approx
  \langle\gamma\rangle \left(
    1+f_\beta\langle \kappa\rangle
  \right)
  \approx
  \frac{\langle\gamma\rangle}{1-f_\beta\langle\kappa\rangle},
\end{equation}
where $f_\beta\equiv \langle \beta^2\rangle/\langle \beta\rangle^2$ is a
redshift-moment ratio of the order of unity
\nocite{1997A&A...318..687S}({Seitz} \& {Schneider} 1997).

Since the tangential distortion signal, $g_T$, is a function of
cluster radius, we first decompose the tangential distortion profile
of our background sample (B), defined above as our reference (see
\S~\ref{sec:samples}), into the following form:
\begin{equation}\label{eq:gtfit_back}
  g_{T,{\rm B}}(\theta) = a_{\rm B}\theta^{-b_{\rm B}},
\end{equation}
where the radial shape of $g_T(\theta)$ is assumed to be a single
power-law with a power index $b_{\rm B}$, and $a_{\rm B}$ represents
the distortion amplitude of our reference background.  In practice, we
fit the outer profile of $g_{T}(\theta)$, excluding the nonlinear
regime ($\theta\lesssim\!1$), to the power-law model, constraining
simultaneously the distortion amplitude $a_{\rm B}$ and the outer
slope, $b_{\rm B}$.  Next, for each of our other defined samples, we
fit a power law with the same slope $b_{\rm B}$, but allow the
amplitude $a_i$ to vary:
\begin{equation}\label{eq:gtfit_i}
  g_{T,i}(\theta) = a_i\theta^{-b_{\rm B}}.
\end{equation}
Therefore, if we calculate the lensing signal of $i$-th sample relative
to the reference background (B),
\begin{equation}\label{eq:gt_ampn}
  \Gamma_i\equiv g_{T,i}(\theta)/g_{T,{\rm B}}(\theta)= a_i/a_{\rm B}.
\end{equation}
From equation~(\ref{eq:gt-d}), we obtain the following expression in the
WL approximation ($|\langle \kappa\rangle|,|\langle
\gamma\rangle|\ll 1$):
\begin{equation}\label{eq:amp_D}
  \Gamma_i=
  a_i/a_{\rm B} \approx \langle \beta \rangle_i/ \langle \beta\rangle_{\rm B} =
  \langle D \rangle_i/\langle D\rangle_{\rm B},
\end{equation}
where $\langle~\rangle_i$ ($i=1,2,...,{\rm B}$)
represents averaging over the redshift distribution $N_i(z_s)$ of
$i$-th galaxy sample. The relative distortion strength $\Gamma_i$ can
be regarded as a function of the discrete background sample $i$ with
the redshift distribution $N_i(z_s)$, which is observationally
available and calibrated by deep, multi-band blank surveys such as the
COSMOS survey.  For a given cosmological model, one can readily
construct its theoretical prediction $\Gamma_i$ ($i=1,2,...,{\rm B})$
using a set of redshift distribution functions $N_i(z_s)$.  The
function $\Gamma_i$ can be formally labeled by its mean redshift
\begin{equation}\label{eq:z_mean}
  \langle z_s\rangle_i\equiv
  \int\!dz_s\,N_i(z_s)\,z_s/\int\!dz_s\,N_i(z_s)
\end{equation}
which is independent of the cosmological model.

To the next order of approximation, the distortion amplitude ratio is
written as \nocite{2007ApJ...663..717M}(Appendix B of  {Medezinski} {et~al.} 2007)
\begin{equation}
  \Gamma_i=
  \frac{\langle D\rangle_i}{\langle D\rangle_{\rm B}}
  \left\{
    1+\left( f_{\beta,i} \langle \beta\rangle_i -
      f_{\beta,{\rm B}}\langle \beta\rangle_ {\rm B}\right)\kappa_\infty(\theta)
    + O(\kappa_\infty^2)
  \right\}
\end{equation}
with $f_{\beta,i}\equiv \langle \beta^2\rangle_i/\langle \beta\rangle_i^2$ and
$f_{\beta,{\rm B}}\equiv \langle \beta^2\rangle_{\rm B}/\langle
\beta\rangle_{\rm
  B}^2$. The next order correction term is proportional to $(f_{\beta,i}
\langle \beta\rangle_i - f_{\beta,{\rm B}}\langle \beta\rangle_ {\rm
  B})\kappa_\infty(\theta)$, which is much smaller than unity for the
galaxy samples of our interest in the mildly nonlinear regime
($\theta\gtrsim 1$).  We thus simply adopt equation (\ref{eq:amp_D})
obtained in the WL approximation. This can be further
justified by the fact that the slope parameter $b_{\rm B}$ is
constrained by a least $\chi^2$ fit to the outer distortion profile.
The mean weighted cluster radius $\langle \theta\rangle\equiv \sum_i
u_{g,i}\theta_i/\sum_i u_{g,i}$ used for a fit is $\langle
\theta\rangle\sim 10-11$\,arcmin for our clusters, where the weak
lensing approximation is valid.

\section{Results}\label{sec:results}

\subsection{Weak lensing profiles}\label{sec:wl-prof}

\begin{figure}
  \centering
   \begin{overpic}[width=8cm,height=5cm]{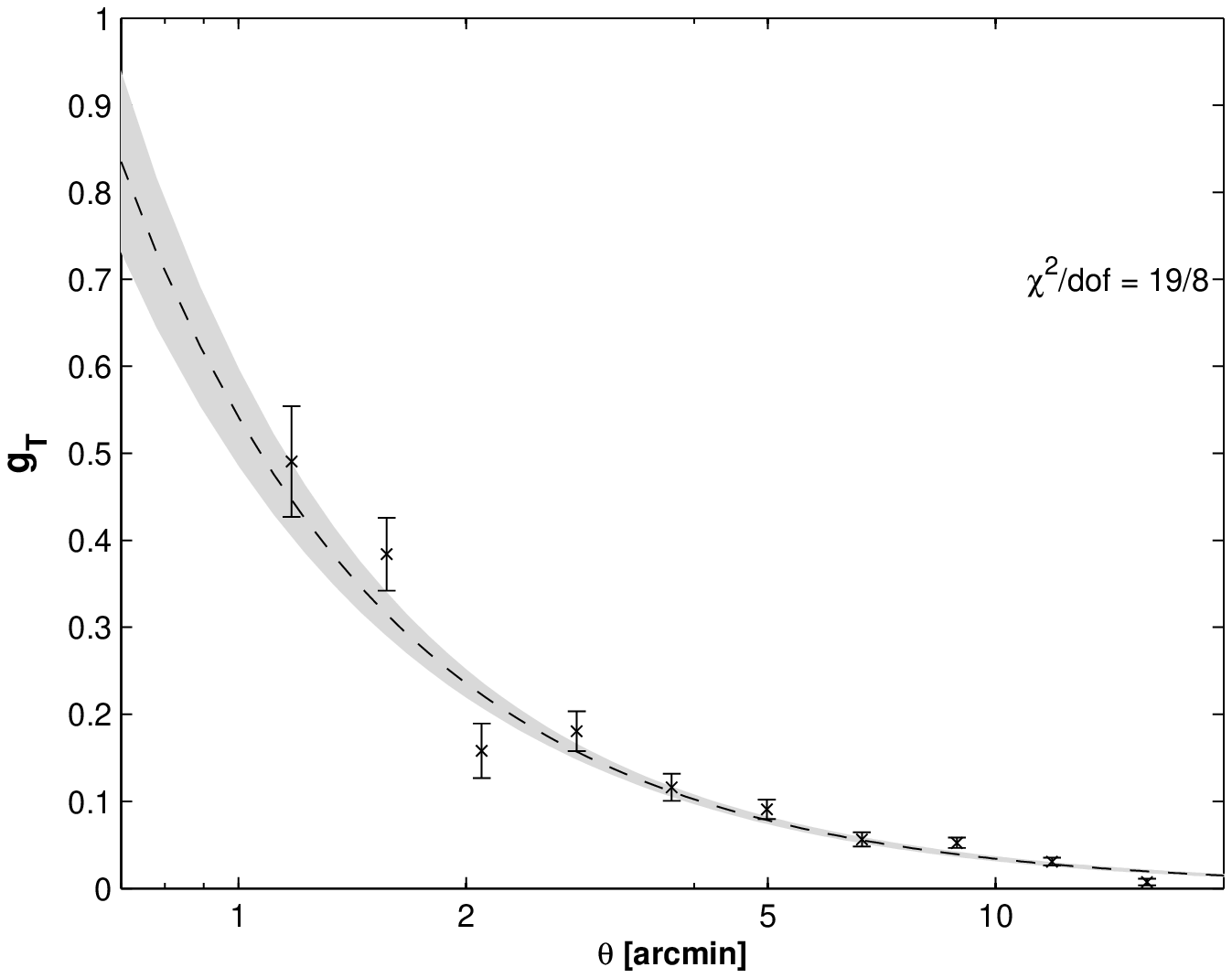}
     \put(47,59){{\bf A370}}\end{overpic}
   \begin{overpic}[width=8cm,height=5cm]{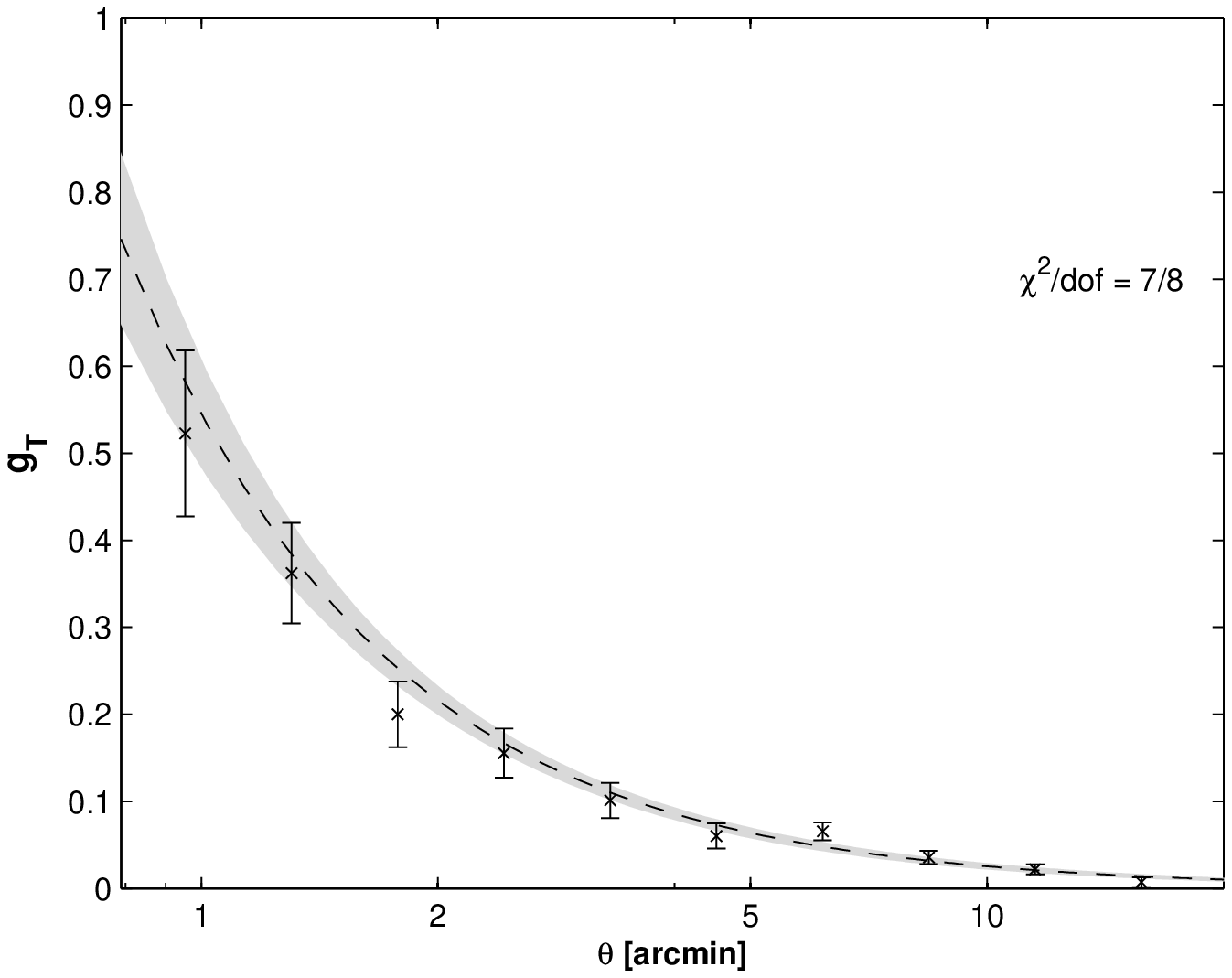}
     \put(45,59){{\bf ZwCl0024}}\end{overpic}
   \begin{overpic}[width=8cm,height=5cm]{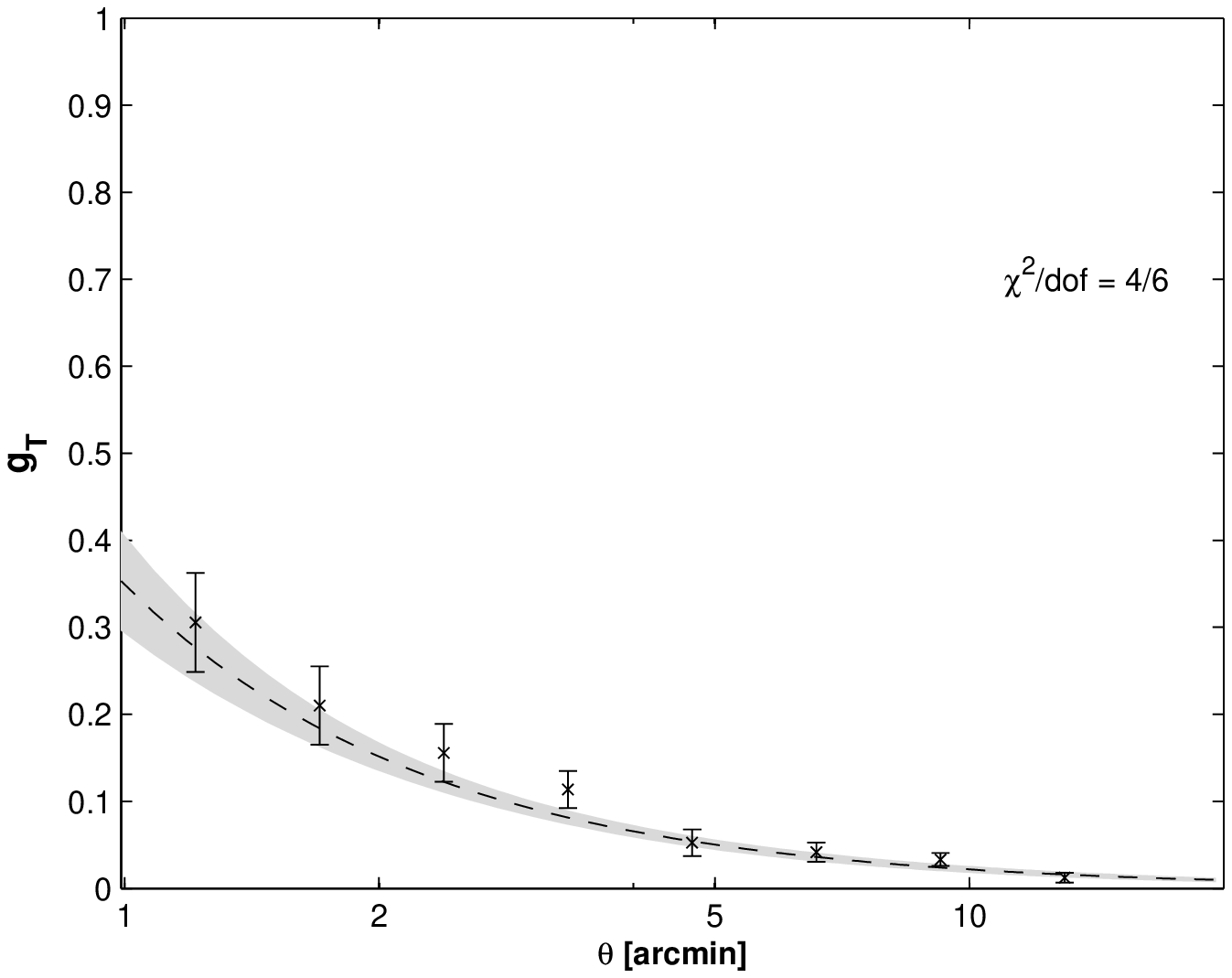}
     \put(45,59){{\bf RXJ1347}}\end{overpic}
   \caption{Tangential distortion $g_T$ vs. distance from cluster
     center for A370 (top), ZwCl0024+17 (middle) and RXJ1347-11
     (bottom) reference background samples. Overlaid is the power-law
     fit (dashed black line) with $1-\sigma$ confidence levels (shaded
     region) and the PL-fit $\chi^2$ is indicated.}
  \label{fig:gt_r_back}
\end{figure}

For each cluster, a reference background sample has been defined
(explained above in \S~\ref{sec:samples}). The WL tangential
distortion, $g_T$, vs. distance from cluster center, $\theta$, of each
reference background is plotted in Fig.~\ref{fig:gt_r_back} (black
crosses). Each reference background sample is fitted by a power-law
according to Eq.~\ref{eq:gtfit_back}, but only in the WL regime, i.e.,
outside $\theta\gtrsim1\arcmin$. The fit is estimated using two ways
-- we fit the entire sample dataset, weighting each galaxy $g_{T,i}$
by $u_{g,i}$ (see \S~\ref{sec:wl}), and we also fit the binned $g_T$
profile, $\langle g_{T}(\theta_n)\rangle$, weighting by the bin error,
$1/{\sigma_{T}}^2(\theta_n)$. We find good consistency between the two
fitting schemes.  The goodness-of-fit $\chi^2$ values of the binned
fit are displayed next to each power-law fit, and also in
Table~\ref{tab:samples}.  As can be seen, a simple power-law serves as
a reasonable fit in all cases.

For each of our defined samples (foreground, orange, green, red, blue and
dropouts) we again plot $g_T$ vs. radius in Figure~\ref{fig:gt_r_samples} (gray,
orange, green, red, blue and dropouts, respectively, top to bottom panels of
each cluster). We fit each sample with the same power-law index, $b_{\rm B}$,
given by its relevant reference background (the sample fit is shown as dashed
black line with $1-\sigma$ confidence bounds, and the reference background fit
is also shown as a dotted line). The WL amplitude of each sample relative to the
reference background is given in each panel as $\Gamma_i\equiv a_i/a_{\rm B}$
(as defined by Eq.~\ref{eq:gt_ampn}) and detailed in Table~\ref{tab:samples}.
We see that indeed, for the foreground samples, the $g_T(\theta)$ profile agrees
with zero throughout, and gives a relative WL amplitude of zero.

\begin{figure}
  \centering
\includegraphics[width=8cm,height=6cm]{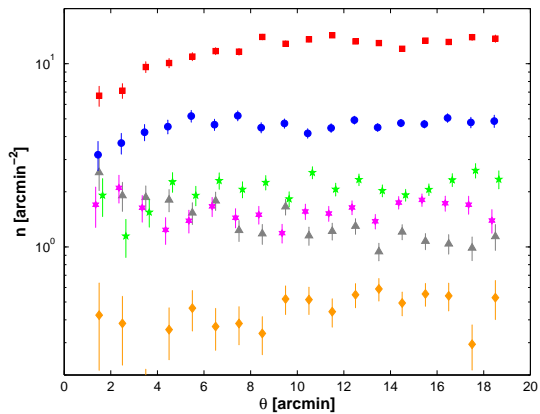}
  \caption{Galaxy surface number density vs. radius for A370 foreground sample
(gray triangles), orange (orange diamonds), green (green pentagrams), red (red
squares), blue (blue circles) and dropout (magenta hexagrams) background samples.}
  \label{fig:dens_r}
\end{figure}

\begin{figure*}
  \centering
   \begin{overpic}[width=8cm]{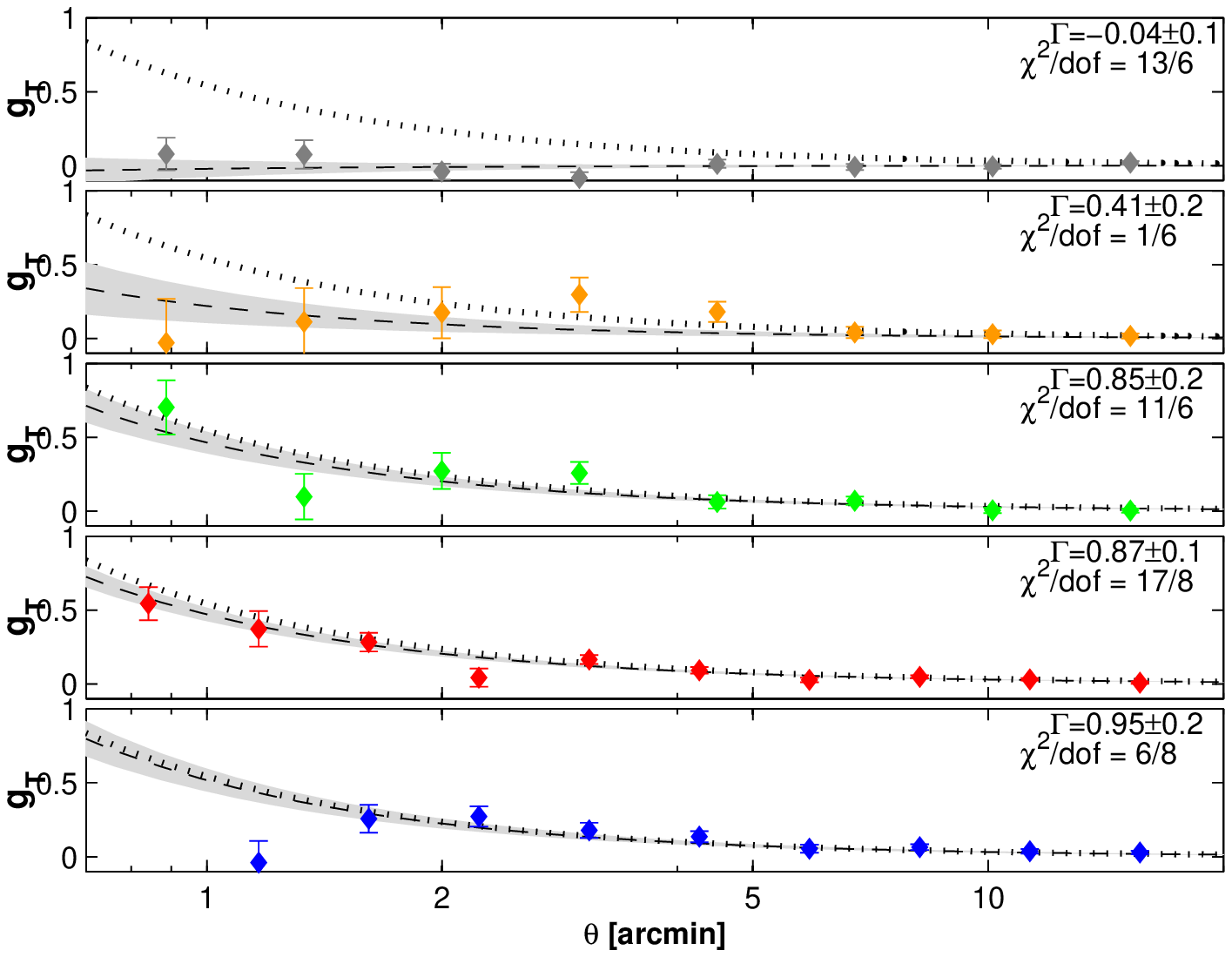}
     \put(47,72){{\bf A370 bright}}\end{overpic}
   \begin{overpic}[width=8cm]{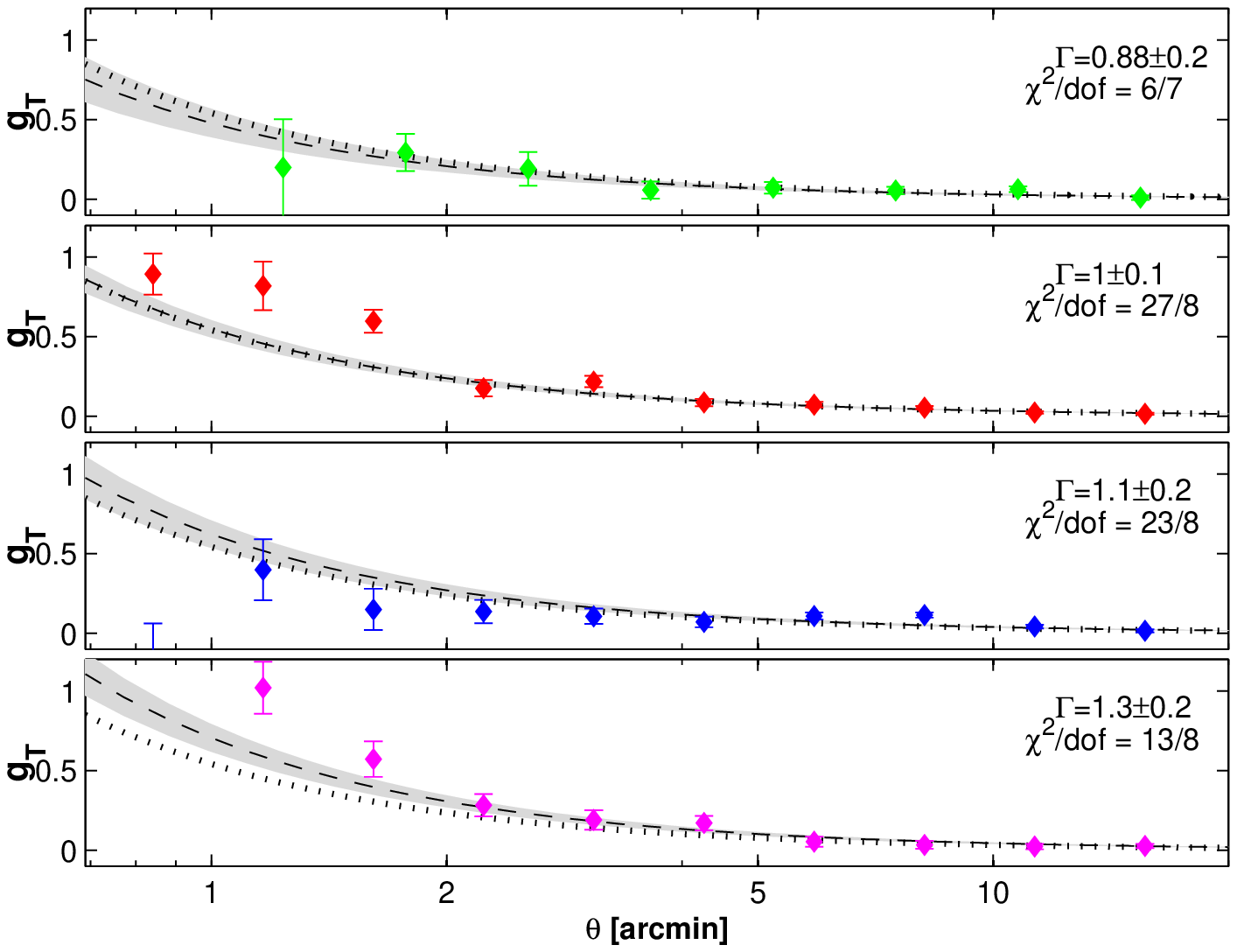}
     \put(47,72){{\bf A370 faint}}\end{overpic}
   \begin{overpic}[width=8cm,height=7cm]{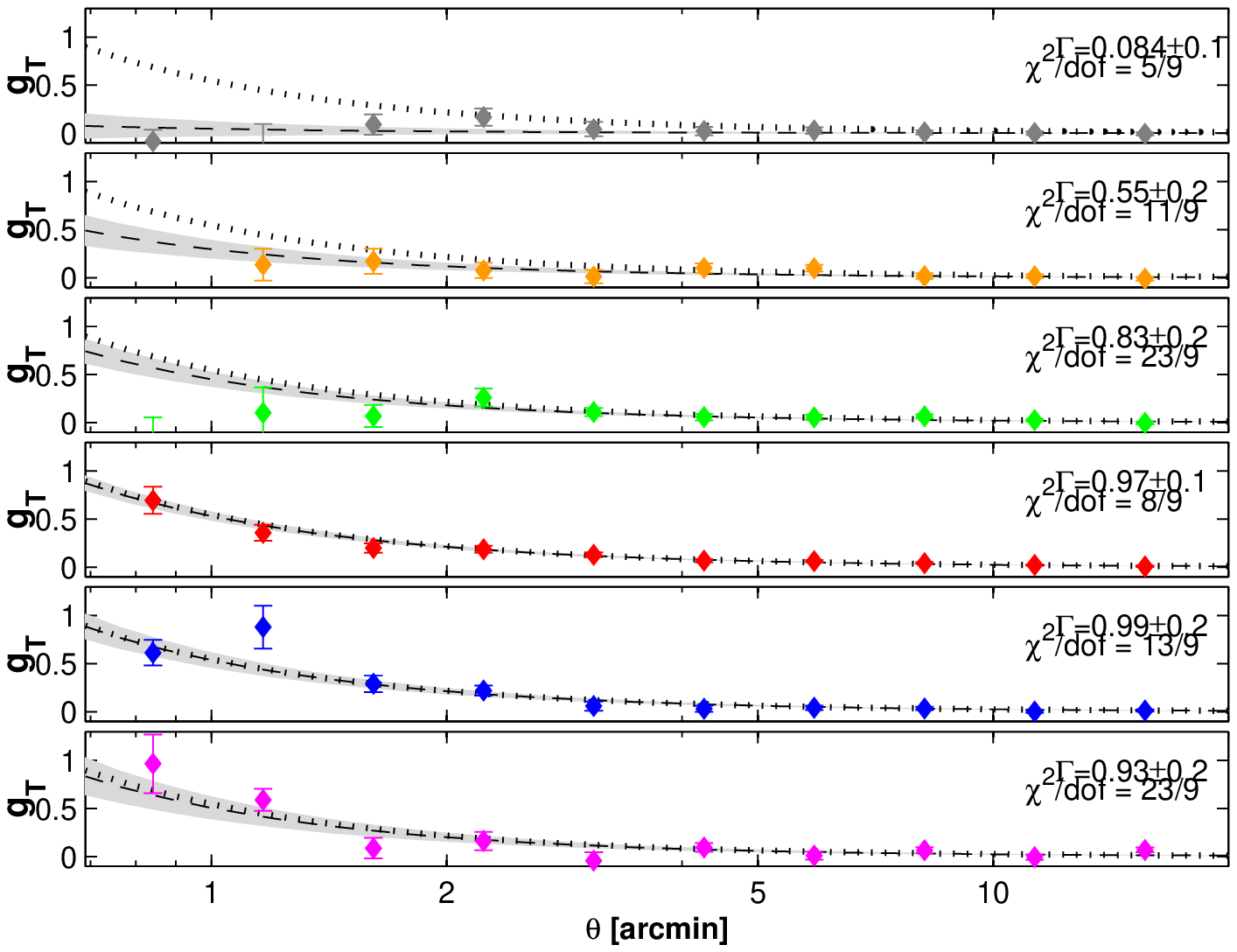}
     \put(47,83){{\bf ZwCl0024}}\end{overpic}
   \begin{overpic}[width=8cm,height=7cm]{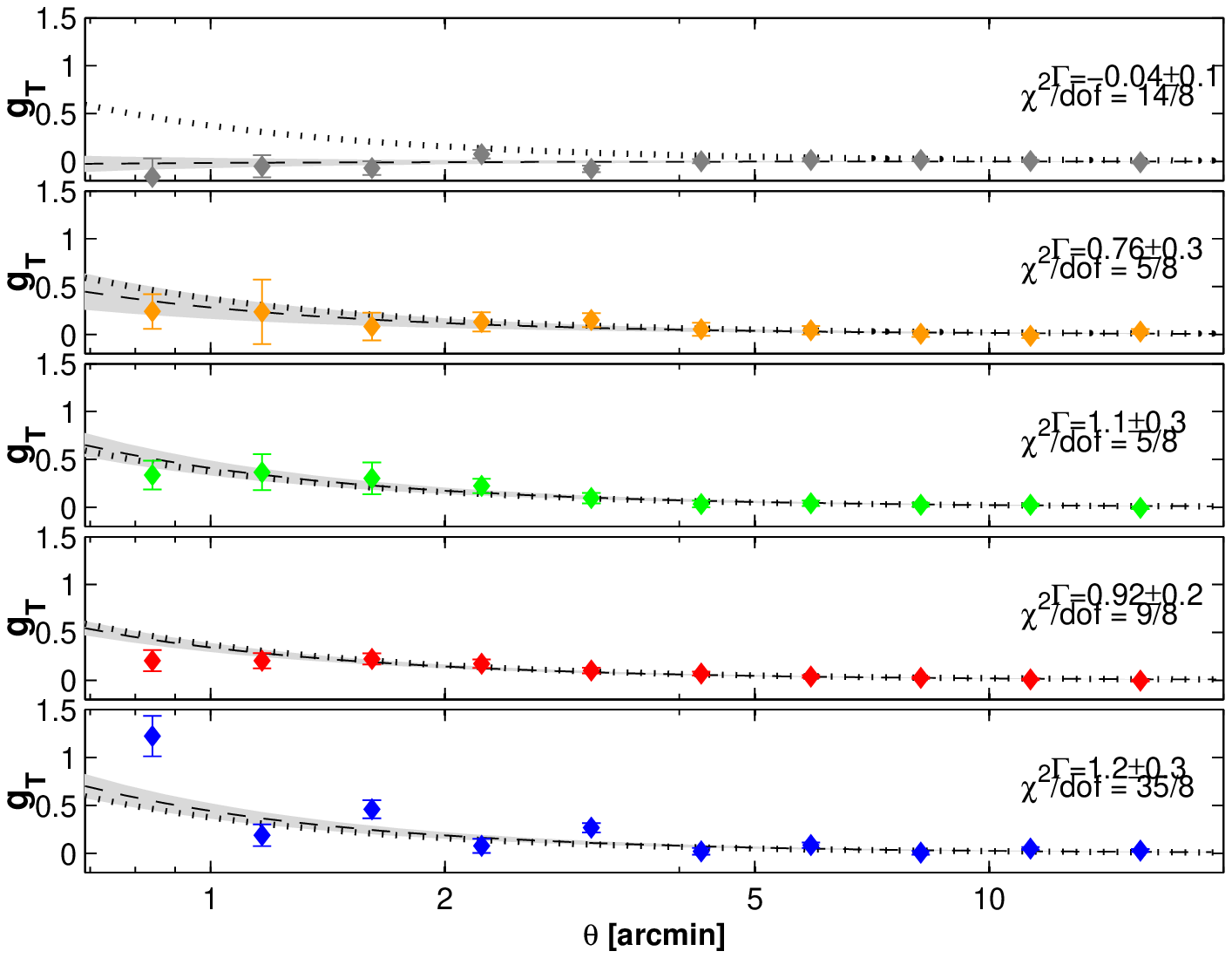}
     \put(47,83){{\bf RXJ1347}}\end{overpic}
   \caption{$g_T$ vs. cluster radius for A370 bright samples (top left:
foreground, orange, green, red \& blue) and faint (top right: green, red, blue
\& dropouts), ZwCl0024+17 (bottom left: foreground, orange, green, red, blue \&
dropouts) and RXJ1347-11 (bottom right: foreground, orange, green, red \& blue),
where the fixed power-law fit is overlaid (dashed black line) with
$1-\sigma$ confidence levels (shaded region). Also plotted is the power-law fit
of the equivalent ``reference'' background sample (dotted curve). In each case
the resulting normalized $g_T$ amplitude ratio, $\Gamma$, is denoted next to the
profile.}
  \label{fig:gt_r_samples}
\end{figure*}

As another consistency check, we plot the galaxy surface number density vs.
radius for A370 samples in Figure ~\ref{fig:dens_r}. As can be seen, no
clustering is observed toward the center for any of the samples, which
demonstrate that there is no contamination by cluster members in the samples
comprising only background members. The foreground sample (gray triangles) shows
a modest increase in number density (factor of 2 increase from $\theta=20$ to
$\theta=2$) compared to the cluster, despite the exclusion of the cluster
early-type galaxies by colour. Bluer later-type cluster members are to be
expected here given the redshift window sampled by reference to
Figure~\ref{fig:all_zhist}, which shows that the tail of the distribution
reaches just beyond the redshift of A370 \S~\ref{sec:cosmos-z}).

The data for A370 represents the best available data, both in terms of
the filter coverage -- $B,R,z'$, and in terms of total exposure times
in each band. For RXJ1347-11 we only have $V,R,z'$ coverage, making it
somewhat harder to set apart the different populations, and also
dropping out between $V$ and $R$ is not as clear since the two filters
are less well separated in wavelength, preventing us from selecting a
dropout sample. For ZwCl0024+17 we do have $B,R,z'$, but the $z'$ band
is much shallower.

\subsection{COSMOS photometric redshifts}\label{sec:cosmos-z}

\begin{figure*}
  \centering
\includegraphics[width=8cm,height=6cm]{COSMOS/COSMOS_A370_colorcolor_bg_100820.eps3}
\includegraphics[width=8.5cm,height=6cm]{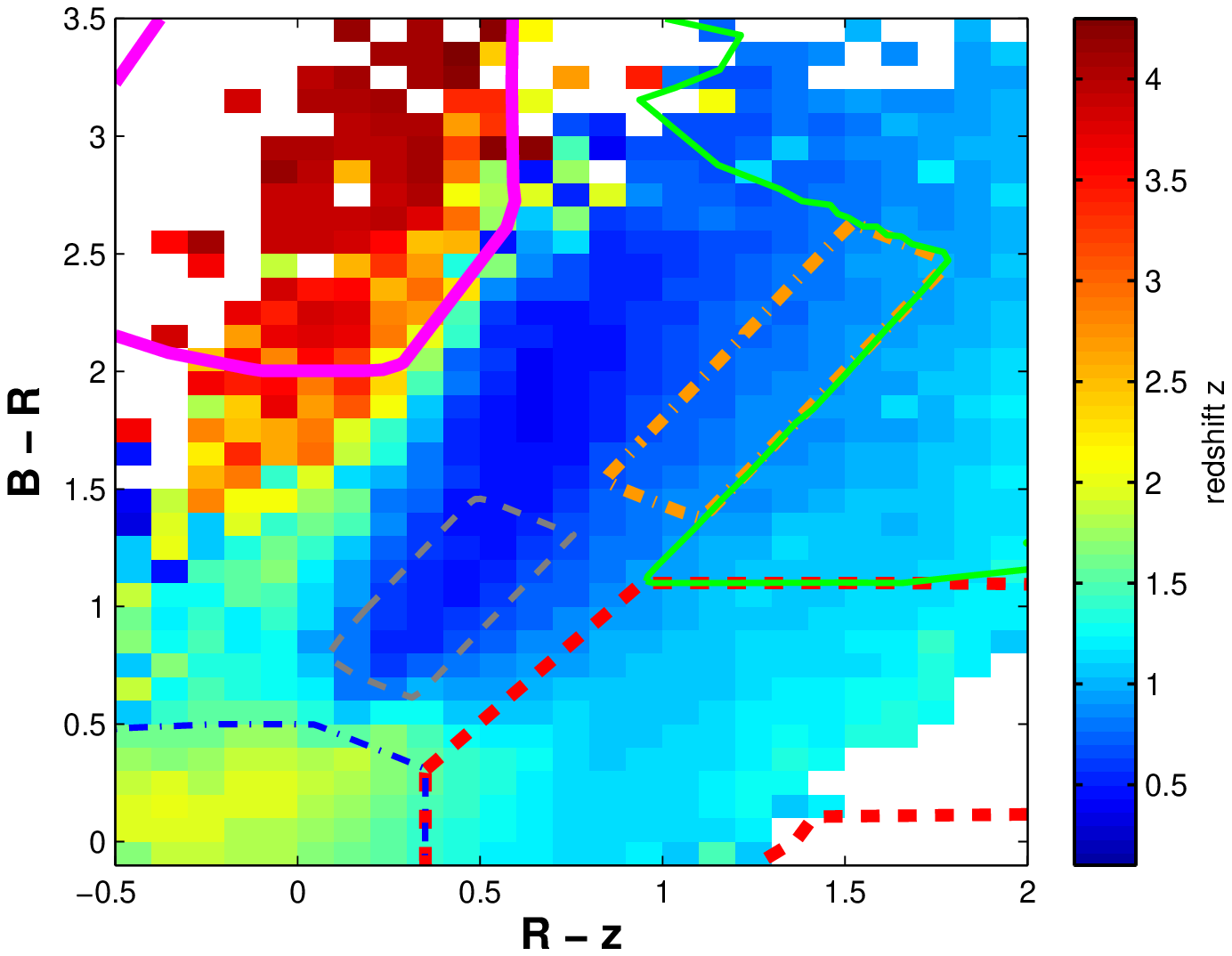}
  \caption{{\it Left:} $B_{\rm J}-R_{\rm C}$ vs. $R_{\rm C}-z'$ CC diagram,
showing the distribution of galaxies in COSMOS field. Applying the same CC-cuts
as our selection for A370, we define galaxy samples: foreground (gray), orange
(orange), green (green), red (red), blue (blue), and dropout (magenta) galaxies,
in order to estimate the mean redshifts and mean depths of these samples using
COSMOS photo-$z$'s. {\it Right:} Average COSMOS redshift in CC bins. Overlaid
are the boundaries of the foreground (thin dashed gray line), orange (thick
dotted-dashed orange line), green (thin solid green line), red (thick dashed red
line), blue (thin dotted-dashed blue line) and dropout (thick solid magenta
line) samples selected in the COSMOS field according to A370 CC-cuts. Evidently,
different regions of CC space correspond to different redshift populations of
galaxies. Most notably, The top left corner denotes dropout galaxies of
$z\gtrsim3.5$, corresponding well to our selection of highly distorted dropout
galaxies.}
  \label{fig:z_CC}
\end{figure*}

To estimate the respective depths of the different samples defined
above from our Subaru photometry, we make use of the accurate
photometric redshifts derived for the well studied multi-band field
survey, COSMOS \nocite{2007ApJS..172...99C}({Capak} {et~al.} 2007).  For COSMOS, photometric
redshifts have been derived by \nocite{2009ApJ...690.1236I}{Ilbert} {et~al.} (2009) using 30
bands in the UV to mid-IR.  Since the COSMOS photometry does not cover
the Subaru $R_{\rm C}$ band, we estimate $R_{\rm C}$-band magnitudes
for it.  For this we use the HyperZ \nocite{2000A&A...363..476B}({Bolzonella}, {Miralles} \&  {Pell{\'o}} 2000)
template fitting code to obtain the best-fitting spectral template for
each galaxy, from which the $R_{\rm C}$ magnitude is derived with the
transmission curve of the Subaru $R_{\rm C}$-band filter
\nocite{2010ApJ...714.1470U}(see {Umetsu} {et~al.} 2010).

\begin{figure}
  \centering
\includegraphics[width=8cm,height=6cm]{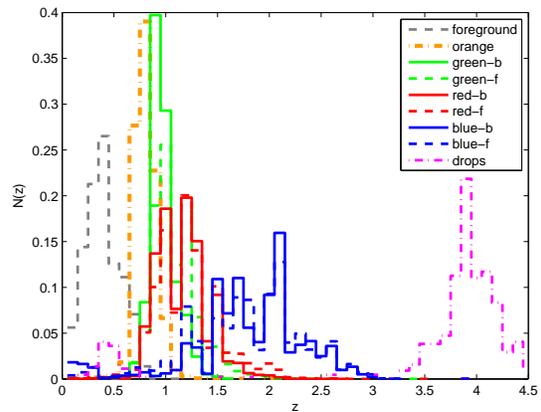}
\caption{Redshift distribution of all A370 samples: foreground (dashed gray),
orange (dotted-dashed orange), green (green) bright (solid) and faint (dashed),
red (red) bright (solid) and faint (dashed), blue (blue) bright (solid) and
faint (dashed), and dropout (dotted-dashed magenta) using COSMOS photo-z's.}
  \label{fig:all_zhist}
\end{figure}

We then select samples by applying the same CC/magnitude limits as we
did above for each of our clusters - A370, ZwCl0024+17 and RXJ1347-11. This
is shown in Fig.~\ref{fig:z_CC} (left) for the COSMOS catalogue and
plotted in terms of the same CC plane as A370. The colour distribution
of COSMOS field galaxies seen in this $B,R,z'$ CC-space is very
similar to that of A370, displaying the same morphology, including
red, blue and dropout populations, but without the density peak
associated with the massive cluster A370. We also show how redshift
varies in this CC-plane by calculating the mean photo-$z$ redshift from
COSMOS in fine bins over the CC plane (Fig.~\ref{fig:z_CC}, right),
with the samples boundaries displayed as well. This demonstrates that
the main overdensity in the CC plane near $B_{\rm J}-R_{\rm C}\sim1$
and $R_{\rm C}-z'\sim0.3$ has a mean redshift of around
$z\lesssim0.5$, which agrees with our estimation for A370 where we
found very little WL signal implying these object lie predominantly in
the foreground of the cluster. We also see from this figure that the
region where we picked ``red'' galaxies corresponds to $z\sim1-1.5$,
and the ``blue'' galaxies occupy a region of mean redshift around
$z\sim2$.  Most notably, the top left corner of ``dropout'' galaxies
corresponds to high-$z$ with $z\gtrsim3.5$.
We further plot the redshift distribution of all the samples in
Fig.~\ref{fig:all_zhist}. We calculate the median redshift of each sample and
summarized in Table~\ref{tab:samples}.

\begin{figure}
  \centering
\includegraphics[width=8cm,height=6cm]{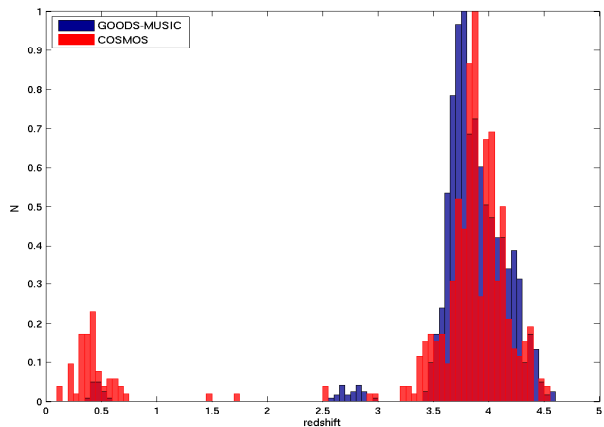}
\caption{Redshift distribution of the dropout-selected sample using
  COSMOS photo-z's (light red) and using GOODS-MUSIC photo-z's (dark
  blue). The low-$z$ peak seen more notably from the COSMOS photo-z's
  is most likely due to misclassified galaxy redshifts, supported by
  the smaller numbers found when using the GOODS-MUSIC photo-z
  catalogue.}
  \label{fig:drops_zhist}
\end{figure}

However, if we look at the distribution of COSMOS redshifts of the
dropout sample (Fig.~\ref{fig:drops_zhist}, red), we find it is somewhat
double-peaked, with most galaxies lying around $z\sim3.5$, but a
significant fraction identified as having $z\sim0.4$. Since we are
certain most of the galaxies in this region are in fact high redshift
dropout galaxies, justified by the apparent high WL distortions
measured for these galaxies (see Fig.~\ref{fig:gt_r_samples}) and by
reference to deep spectroscopic work in this rising plume of
``dropout'' galaxies \nocite{1999ApJ...519....1S}({Steidel} {et~al.} 1999), we may conclude the
low redshifts assigned to some of these galaxies may be misclassified
as low redshift early type/dusty galaxies in the photo-$z$ catalogue of
the COSMOS field. 
This is not surprising, since at faint magnitudes the COSMOS photo-z
have a relative high catastrophic failure rate
\nocite{2009ApJ...690.1236I}({Ilbert} {et~al.} 2009).

We further examine this issue using the somewhat deeper GOODS-MUSIC
catalogue \nocite{2006A&A...449..951G,2009A&A...504..751S}({Grazian} {et~al.} 2006; {Santini} {et~al.} 2009), which has 15
bands, including high quality ACS photometry (GOODS-S) and deep IRAC
imaging which is very helpful in reducing the outlier rate a high-$z$.
Using the $z'$-band selected GOODS-MUSIC photo-z, we derive a spectral
classification for each galaxy using a library of $\sim200$ PEGASE
\nocite{1997A&A...326..950F}({Fioc} \& {Rocca-Volmerange} 1997) templates very similar to that described
in \nocite{2006A&A...449..951G}{Grazian} {et~al.} (2006), as included in the EAZY software
\nocite{2009ApJ...706L.173B}({Brammer} {et~al.} 2009), and then calculate Subaru magnitudes for
all the galaxies using the BPZ \nocite{2000ApJ...536..571B}({Ben{\'{\i}}tez} 2000) code.

By making the same CC selection, we plot the redshift distribution of
the same dropout sample (Fig.~\ref{fig:drops_zhist}, blue), but here
practically no low-$z$ peak is observed, and all galaxies in this
region are estimated to have high redshifts, $z\gtrsim2.5$. This
comparison between COSMOS and GOODS-MUSIC redshifts allows us to
securely set a conservative lower $B-R$ limit to avoid inclusion of
real low-$z$ objects. We can thus safely assume all objects identified
as low-$z$ in the COSMOS catalogue are largely mistakenly classified,
a point also made in relation to this by \nocite{2010A&A...516A..63S}{Schrabback} {et~al.} (2010).
Estimating the median redshift of the
  dropout sample gives $z\simeq3.8$, a value very close to the mean
  redshift of all galaxies lying above $z>1.5$ galaxies. This further
  demonstrates the low-$z$ peak is a low-significance contamination.
A further examination of
the photometric redshift estimation for such objects in the COSMOS
field seems worthwhile in view of these results.

\subsection{Lensing strength dependence on magnitude}\label{sec:gt-mag}

\begin{figure}
  \centering
  \includegraphics[scale=0.5]{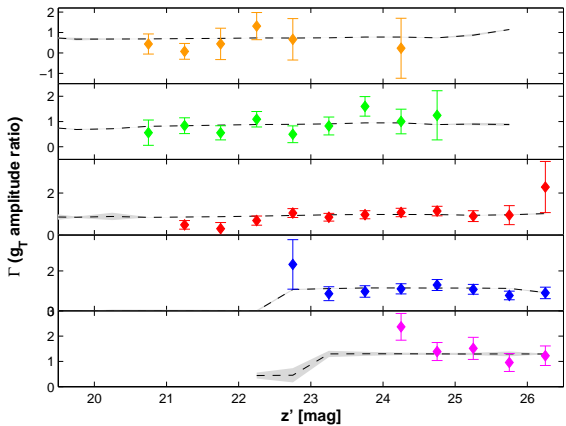}
  \caption{$\Gamma$, the $g_T$ amplitude ratio, vs. magnitude ($z'$
    band) for A370 orange, green, red, blue and dropout galaxy samples (top to
    bottom). Overlaid is the lensing depth, $\dlsds$, vs. magnitude
    calculated from COSMOS photo-z's for each of the samples (dashed
    black line) with $1-\sigma$ confidence levels (shaded region).}
  \label{fig:gt_mag}
\end{figure}

As another check, we plot $\Gamma$, the $g_T$-amplitude ratio, vs. $z'$-band
magnitude for A370 selected samples, to see if there is any trend of
the lensing amplitude with magnitude. We also plot the mean $\dlsds$
as a function of magnitude (all values normalized to the reference
background values) calculated in independent magnitude bins for each
sample using the COSMOS photo-$z$ catalogue. This serves as a further
consistency check. Interestingly, for the blue galaxies (and the
dropout galaxies to some extent), the mean signal seems to drop
slightly with fainter magnitudes. This trend is somewhat
counter-intuitive, since we expect that fainter galaxies will be at
higher redshifts, and therefore have on average a higher signal.

The diminished WL signal could possibly hint at a problem with
estimating the WL signal from faint {\bf blue} galaxies, which are in
general quite irregular in morphology, and empirical simulations with
higher space based resolution can be made to examine this better.  The
declining trend of the predicted $\dlsds$ in the case of blue
galaxies, based on the COSMOS photo-z estimation, shown in
Fig.~\ref{fig:gt_mag} (dashed curves), could possibly point to a
miss-classification of blue galaxies with photo-$z$ methods, a well
known problem for blue galaxies, or even simply a limiting magnitude
beyond which the photo-$z$ method fails in this catalogue. We set our
magnitude limits conservatively to minimize these possible problems,
with $19<z'<23$ for the orange sample, $20<z'<24.5$ for the green sample,
$23<z'<25$ for the blue sample, $22<z'<26$ for the red sample, and
$24.5<z'<26.5$ for the dropout
sample, in the case of A370.
Similar examination and limits were also
applied for ZwCl0024+17 and RXJ1347-11, with results discussed below.

\subsection{Lensing strength vs. redshift}\label{sec:gt-z}

\begin{figure}
   \begin{overpic}[width=8cm]{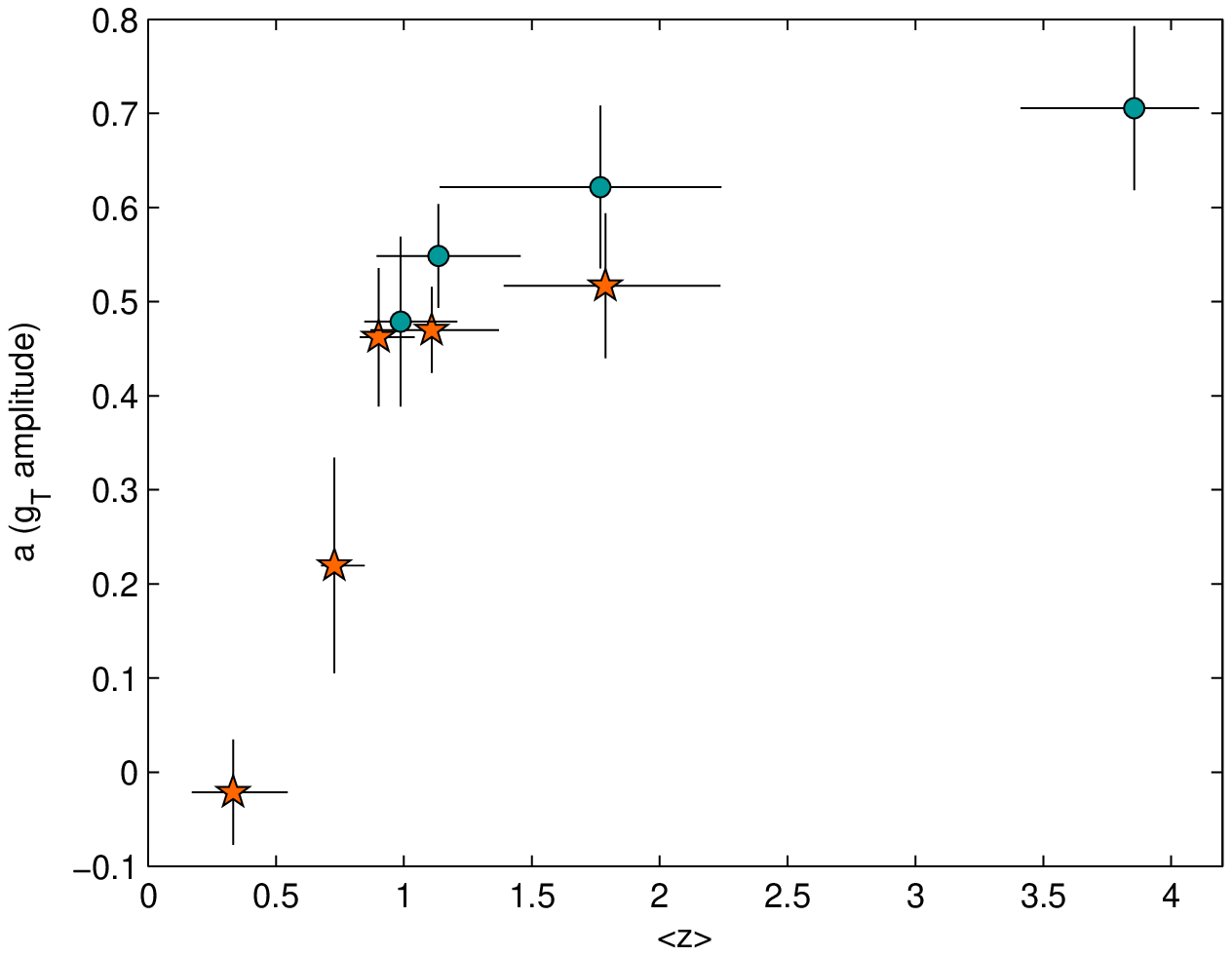}
     \put(47,72){{\bf A370}}\end{overpic}
   \begin{overpic}[width=8cm]{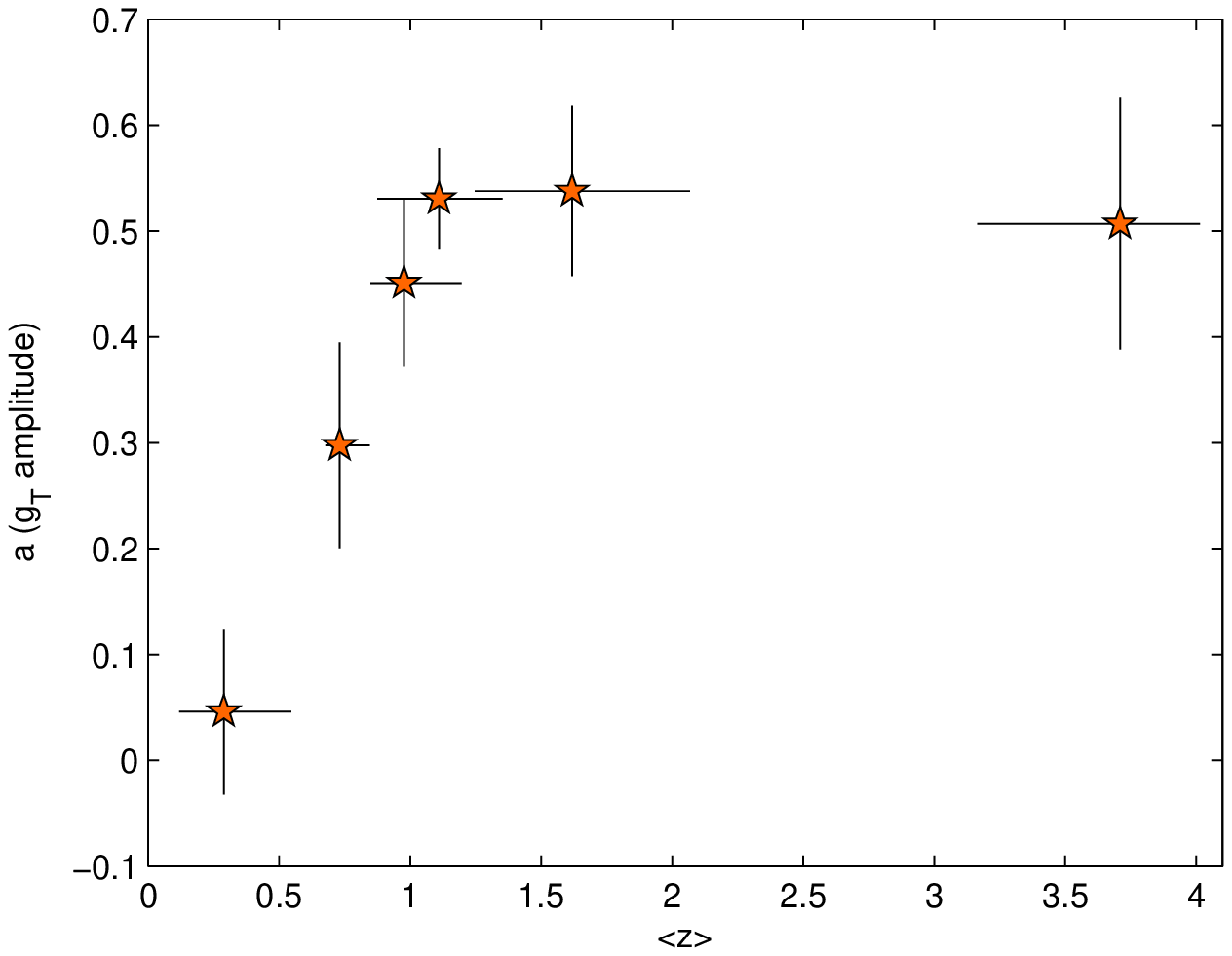}
     \put(45,72){{\bf ZwCl0024}}\end{overpic}
   \begin{overpic}[width=8cm]{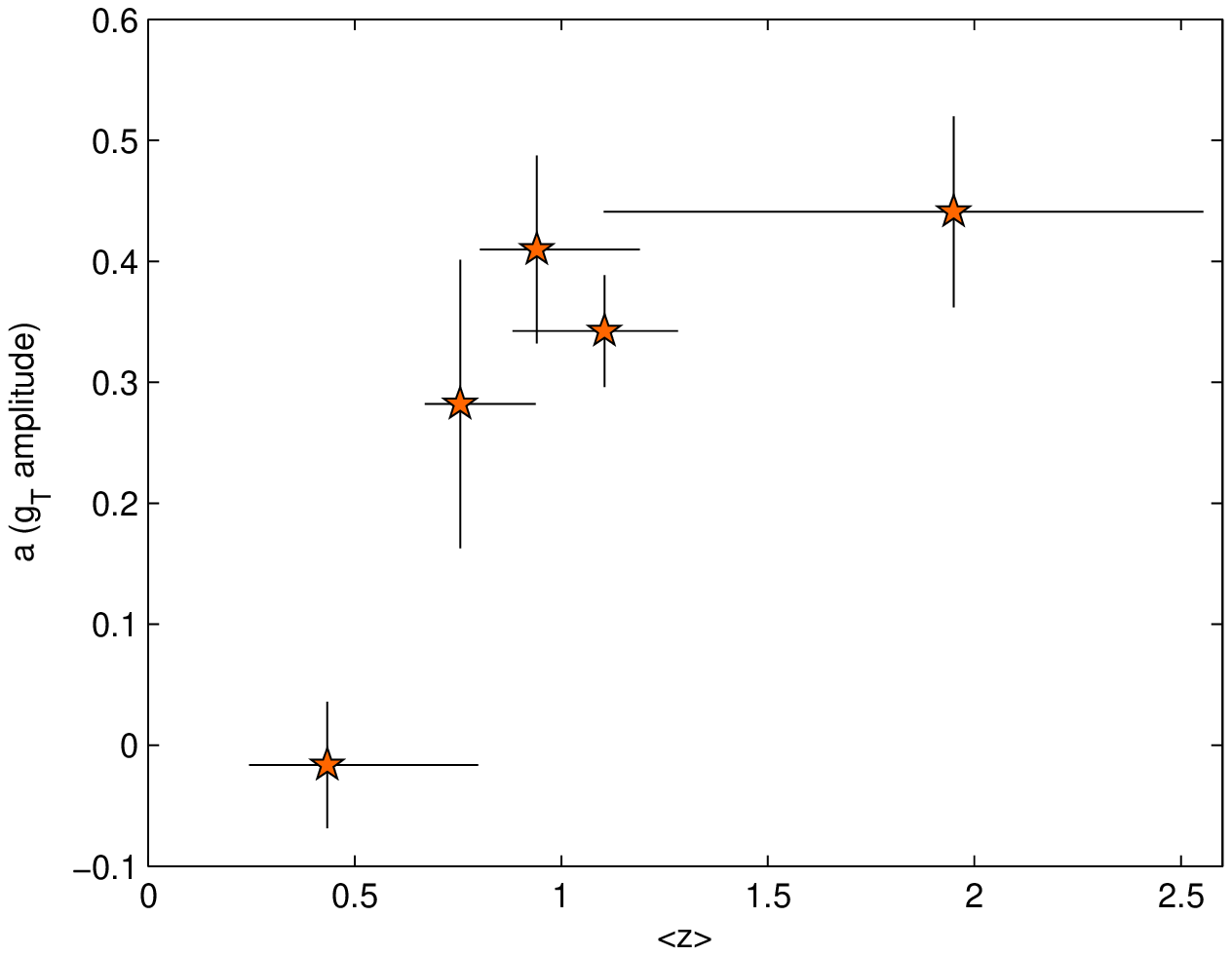}
     \put(45,72){{\bf RXJ1347}}\end{overpic}
   \caption{$g_T$ amplitude, $a$, vs. redshift for A370 (top)
     bright (stars) and faint (circles) samples, ZwCl0024+17 (middle)
     and RXJ1347-11 (bottom). A trend of higher WL amplitude
     with redshift is seen. }
  \label{fig:gt_z_unnorm}
\end{figure}
\begin{figure}
   \begin{overpic}[width=8cm]{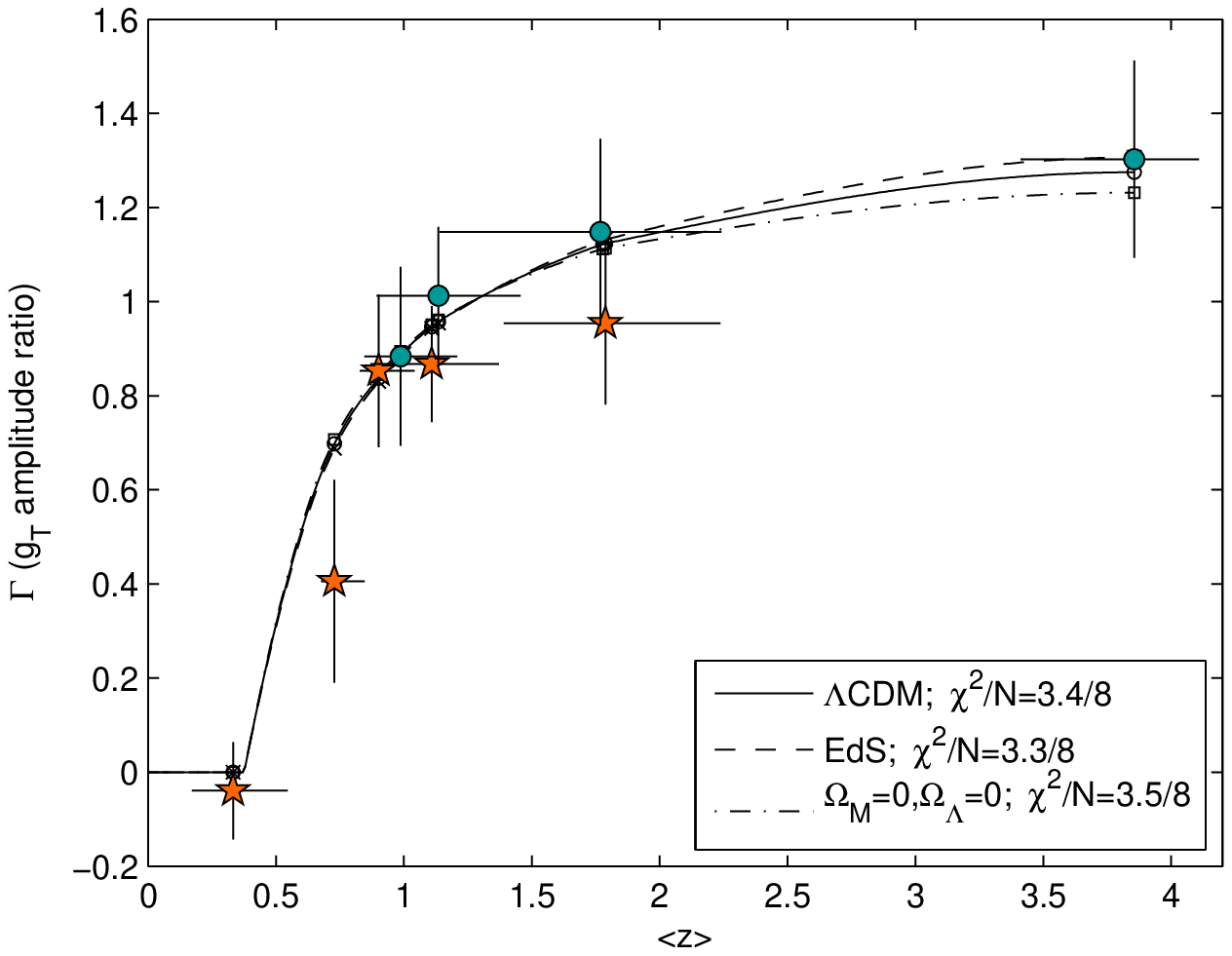}
     \put(47,72){{\bf A370}}\end{overpic}
   \begin{overpic}[width=8cm]{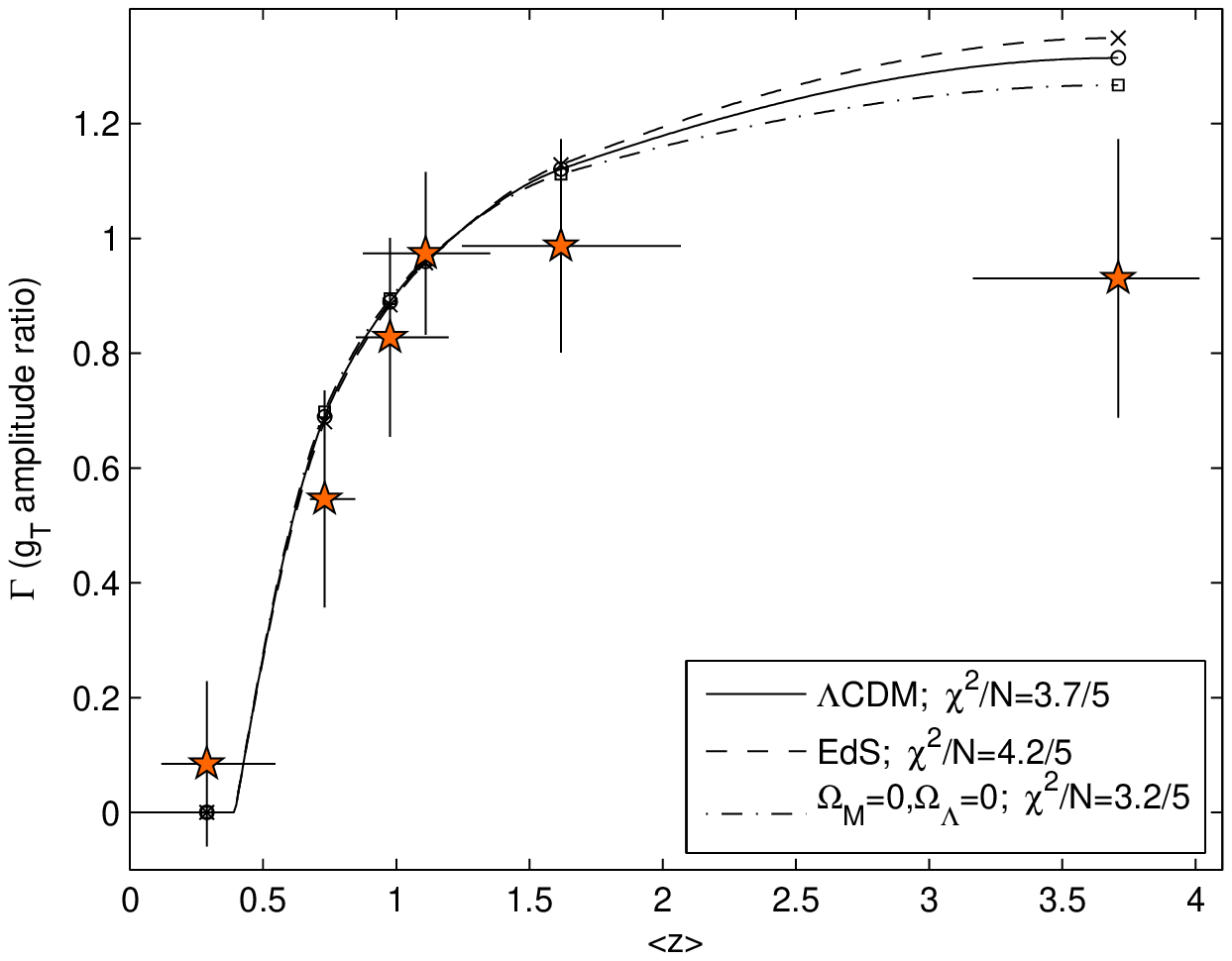}
     \put(45,72){{\bf ZwCl0024}}\end{overpic}
   \begin{overpic}[width=8cm]{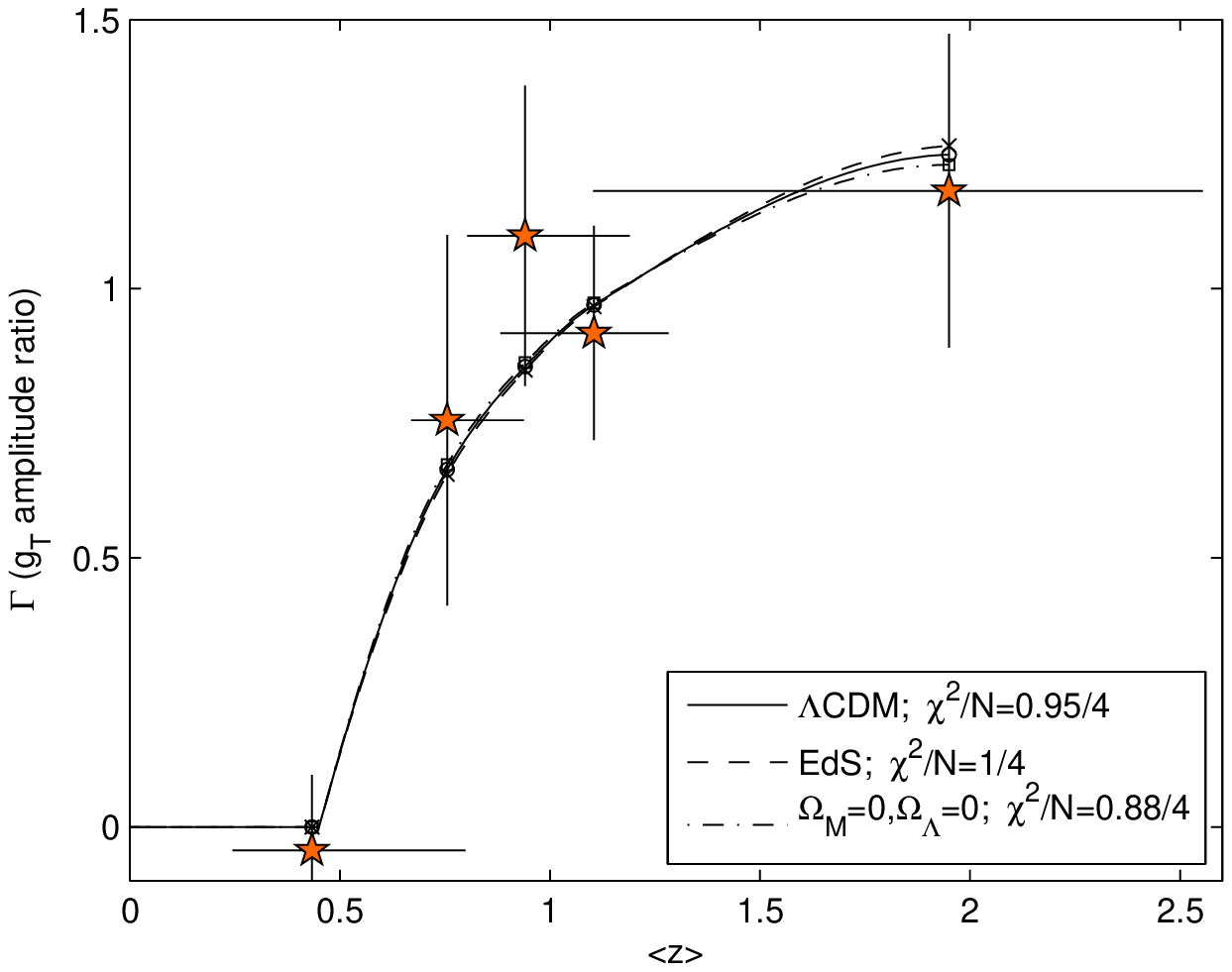}
     \put(45,72){{\bf RXJ1347}}\end{overpic}
   \caption{$g_T$ amplitude ratio, $\Gamma$, vs. redshift for A370 (top)
     bright (stars) and faint (circles) samples, ZwCl0024+17 (middle)
     and RXJ1347-11 (bottom). Also plotted is the lensing depth,
     $\dlsds$, vs.  redshift for different cosmologies - $\Lambda$CDM
     (circles+solid line), Einstein-de Sitter (crosses+dashed line),
     and an empty Universe (squares+dashed-dotted line) estimated
     using the COSMOS photometric redshift catalogue. Horizontal bars represent
the width of the source redshift distribution, given as the 68\% range of data
about the median redshift.}
  \label{fig:gt_z}
\end{figure}

We may now finally combine the WL distortion information with the redshift
information for all the samples of background galaxies defined above.  First, we
plot the WL amplitude, $a$, as a function of redshift in
Fig.\ref{fig:gt_z_unnorm} for the three clusters -- A370 (top), ZwCl0024+17
(middle) and RXJ1347-11 (bottom). A clear trend of increasing WL amplitude is
seen with redshift, especially in the case of A370. A null result is easily
excluded with low significance -- $\chi^2/N=91/8$ for A370, $\chi^2/N=34/4$
for ZwCl0024+17 and $\chi^2/N=42/3$ for RXJ1347-11, for a case of
non-increasing horizontal line.
In order to compare the resulting trend with the cosmological trend,
we plot $\Gamma$, the $g_T$-amplitude ratio, against median redshift,
$\langle z_s\rangle$, in Fig.~\ref{fig:gt_z} for each of the clusters
examined -- A370 (top), ZwCl0024+17 (middle) and RXJ1347-11 (bottom).
The horizontal bars show the width of the source redshift distribution, given as
the 68\% range of data about the median redshift.
Each point in a figure represents an independent sample, where the
first point represents the foreground sample in front of each cluster,
and the other points represent background galaxy samples.  For each
sample, we also calculate the median lensing distance ratio, $\langle
\dlsds\rangle$, for each cosmological model -- $\Lambda$CDM (empty
circles), Einstein-de Sitter (crosses), and an empty universe (empty
squares), using the COSMOS photo-z measurements of galaxies within the
same CC-magnitude boundaries defined for each background sample and
thus obtain the predicted depth. We interpolate between these discrete
predicted values to provide the theoretical relation, $\dlsds(z)$ for
each cosmological model. We can thus compare how the WL strength $\Gamma$
agrees with predicted $\dlsds$. It is quite evident, especially in the
case of A370, where we have the best dataset and therefore more data
points to compare with, that the WL amplitude agrees  well with
the theoretical relations for $\dlsds$ as a function of redshift. For
the other two clusters, ZwCl0024+17 and RXJ1347-11 the data is more
shallow and hence noisier as can be seen in Fig.~\ref{fig:gt_z}
(middle and bottom panels, respectively) but consistency is also found
within the errors. Note that in the case of RXJ1347-11 the colour
separation is not as good, since the bluest band available is $V$,
rather than $B$, limiting the separation in depth of the background
galaxy populations.

\section{Constraining Cosmological Parameters}\label{sec:cosmo}

The clear detection here of the distance-redshift relation from our WL
analysis of A370, prompts the question of how many such clusters would
be required in order to provide a useful cosmological constraint.
This issue has been explored more generally in the context of planned
field and cluster surveys by
\nocite{2003PhRvL..91n1302J,2007MNRAS.374.1377T,2007MNRAS.376..771K}{Jain} \& {Taylor} (2003); {Taylor} {et~al.} (2007); {Kitching} {et~al.} (2007).
\nocite{2007MNRAS.374.1377T}{Taylor} {et~al.} (2007) present a detailed analysis of the
sensitivity of cosmological cluster surveys to the ratio of shear
values measured in independent redshift bins, finding that a large
fraction of the potential integrated signal on the sky is contributed
by abundant small, cluster mass range ($M \approx 10^{14} M_\odot$),
but that a large contribution also comes from the largest clusters,
like those studied here.

Here we use an order-of-magnitude calculation to estimate how well our
newly approved
MCT/CLASH\footnote{\url{http://www.stsci.edu/~postman/CLASH/}} survey
(P.I.  M. Postman) can do in this context, for which we aim to
complete very high quality WL data for approximately 25 massive
clusters, similar in quality to the $B_{\rm J}R_{\rm C}z'$ imaging
of A370.

As defined above in Eq.~(\ref{eq:gt_ampn}), $\Gamma$, is a shear ratio
statistic between any two independent redshift bins summed over
$N_{cl}$ clusters, given by: \be \Gamma_{ij} \approx
\frac{\gamma_i}{\gamma_j}=
\frac{r[\chi(z_j)]r[\chi(z_i)-\chi(z_l)]}{r[\chi(z_i)]r[\chi(z_j)-\chi(z_l)]},
\ee where $r=r(\chi)$ is the comoving angular diameter distance, $\chi(z)$ is
the comoving distance and $z_l$
is the redshift of the lens, and $\Gamma$ scales with the dark energy
equation of state parameter, $w$, as $ \Gamma \approx |w|^{-0.02}$
\nocite{2007MNRAS.374.1377T}({Taylor} {et~al.} 2007).  The fractional error on $w$ is given by
\nocite{2007MNRAS.374.1377T}({Taylor} {et~al.} 2007),
\be 
\frac{\Delta w}{w} = \frac{2}{\gamma_T} \left(\frac{d \ln
    \Gamma}{d \ln w} \right)^{-1} \frac{\sigma_e}{\sqrt{N_b}}, \label{eq:delw}
\ee
where $\gamma_T$ is the typical mean tangential shear of each cluster,
and $\sigma_e=0.3$ is the measured intrinsic scatter in galaxy
ellipticity per mode (KSB), and $N_b$ is the total number of galaxies
summed up behind all the clusters co-added for this purpose.

Assuming $\gamma_T\approx0.05$ and $N_b\approx0.6\times10^6$, summed
over the available background for 25 clusters, (taking A370 as our
guide to the number of background galaxies detected per cluster) we
find from Eq.~(\ref{eq:delw}) the expected precision on $w$ from our sample
is $\Delta w/w\approx0.8$. While this seems a relatively large uncertainty, the
first application of the method by \nocite{2007MNRAS.376..771K}{Kitching} {et~al.} (2007) suggests
that the error distribution is non-Gaussian, with a rather sharp
cut-off at high values of $w$ placing a relatively tight upper limit.
Other geometric probes currently do not succeed much better than this
individually, e.g., from Baryon acoustic oscillations and SN-Ia,
$\Delta w\approx0.3$.  Furthermore, the shear-ratio test has a
different degeneracy with respect to the cosmological parameters to
other probes, making even a crude measurement worthwhile.

This constraint may be improved upon by conducting a more careful
analysis incorporating a proper likelihood function with reasonable
priors for estimating the shear signal and redshift dependence. In addition,
the photometry may be optimised for this purpose with emphasis on
maximising dropout populations so that a wider redshift coverage may be
achieved for better defining the distance-redshift relation.

\section{Discussion \& Conclusions} \label{sec:summary}

Using deep observational data,
the dependence of the amplitude of WL with source distance has been
measured for individual massive galaxy clusters using independent
samples of foreground and background galaxies of differing depths,
with a visible increasing trend.  This is most clearly visible for
A370 where we have Subaru imaging of relatively high quality in $B$,
$R$ and $z'$ bands, allowing us to further subdivide these samples
into independent bright and faint populations.  A general increasing
trend with redshift is also seen for the other two clusters, albeit
noisier than for A370. These clusters datasets are less deep, and with a
reduced colour coverage (only $V,R,z'$ in the case of RXJ1347-11),
demonstrating the advantage of depth and the wide bandwidth
coverage of A370.
Small number statistics and possible dilution, especially in the case
of the dropout samples, may lead to large uncertainties and
underestimated values.  Our photometry comprises only three optical
bands per cluster and so we do not rely on photo-z estimates, but
instead we determine the depth of these background populations with
reference to the very well studied COSMOS and GOODS-MUSIC fields, by
applying CC and magnitude cuts equal to that of our background
populations.

For A370, the trend of increasing WL-amplitude with redshift uncovered here
follows the expected form of the lensing distance-redshift relation but with
uncertainties presently too large to distinguish between cosmologies.  The
encouraging results from the clusters examined here,  most notably for A370,
merit further application of this approach to a larger sample of clusters.

The recently approved MCT/CLASH program will observe 25 clusters with
{\it HST} ACS/WFC3, most of which have deep multi-colour Subaru
imaging, so that we estimate a WL based precision on $w$ of $\Delta
w\approx0.6$, but with a different degeneracy relative to other
probes, complementing existing methods. Combining this WL estimate
with the distance-redshift relation from strong lensing will provide
an enhanced geometric-based cosmological constraint.

To extract the cosmological parameters from such accurate data will
require further refinement of the method. We must take account of
the greater mean depth that lensing magnification generates whose
effect on cluster lensing has been explored in some detail previously
\nocite{1995ApJ...438...49B}({Broadhurst} {et~al.} 1995), and also a second order correction for
the surface density described in \S~\ref{sec:form}. These effects,
although small in terms of the lensing amplitude, are comparable with
the relatively small differences of interest between competing
cosmologies and thus must be explored in any serious study of
cosmology with this method.

\section*{Acknowledgments}
 We are grateful to N.~Kaiser for making the IMCAT
package publicly available. We thank the anonymous referee for
useful comments and important suggestions which greatly improved the
quality of the manuscript.
EM thanks Eran Ofek for his publicly
available Matlab scripts. Work at Tel-Aviv University was supported by
Israel Science Foundation grant 214/02 and supported in part by
National Science Council of Taiwan under the grant
NSC97-2112-M-001-020-MY3.

\bibliography{}


\begin{thebibliography}{}

\bibitem[{Adelman-McCarthy}, {Ag{\"u}eros},  {Allam}, {Allende Prieto}, {Anderson}, {Anderson}, {Annis}, {Bahcall},  {Bailer-Jones}, {Baldry}, {Barentine}, {Bassett}, {Becker}, {Beers}, {Bell},  {Berlind}, {Bernardi}, {Blanton}, {Bochanski}, {Boroski}, {Brinchmann},  {Brinkmann}, {Brunner}, {Budav{\'a}ri}, {Carliles}, {Carr}, {Castander},  {Cinabro}, {Cool}, {Covey}, {Csabai}, {Cunha}, {Davenport}, {Dilday}, {Doi},  {Eisenstein}, {Evans}, {Fan}, {Finkbeiner}, {Friedman}, {Frieman},  {Fukugita}, {G{\"a}nsicke}, {Gates}, {Gillespie}, {Glazebrook}, {Gray},  {Grebel}, {Gunn}, {Gurbani}, {Hall}, {Harding}, {Harvanek}, {Hawley},  {Hayes}, {Heckman}, {Hendry}, {Hindsley}, {Hirata}, {Hogan}, {Hogg}, {Hyde},  {Ichikawa}, {Ivezi{\'c}}, {Jester}, {Johnson}, {Jorgensen}, {Juri{\'c}},  {Kent}, {Kessler}, {Kleinman}, {Knapp}, {Kron}, {Krzesinski}, {Kuropatkin},  {Lamb}, {Lampeitl}, {Lebedeva}, {Lee}, {Leger}, {L{\'e}pine}, {Lima}, {Lin},  {Long}, {Loomis}, {Loveday}, {Lupton}, {Malanushenko}, {Malanushenko},  {Mandelbaum}, {Margon}, {Marriner}, {Mart{\'{\i}}nez-Delgado}, {Matsubara},  {McGehee}, {McKay}, {Meiksin}, {Morrison}, {Munn}, {Nakajima}, {Neilsen},  {Newberg}, {Nichol}, {Nicinski}, {Nieto-Santisteban}, {Nitta}, {Okamura},  {Owen}, {Oyaizu}, {Padmanabhan}, {Pan}, {Park}, {Peoples}, {Pier}, {Pope},  {Purger}, {Raddick}, {Re Fiorentin}, {Richards}, {Richmond}, {Riess}, {Rix},  {Rockosi}, {Sako}, {Schlegel}, {Schneider}, {Schreiber}, {Schwope}, {Seljak},  {Sesar}, {Sheldon}, {Shimasaku}, {Sivarani}, {Smith}, {Snedden}, {Steinmetz},  {Strauss}, {SubbaRao}, {Suto}, {Szalay}, {Szapudi}, {Szkody}, {Tegmark},  {Thakar}, {Tremonti}, {Tucker}, {Uomoto}, {Vanden Berk}, {Vandenberg},  {Vidrih}, {Vogeley}, {Voges}, {Vogt}, {Wadadekar}, {Weinberg}, {West},  {White}, {Wilhite}, {Yanny}, {Yocum}, {York}, {Zehavi}, \&  {Zucker} 2008]{2008ApJS..175..297A}
{Adelman-McCarthy} J.~K., {Ag{\"u}eros} M.~A., {Allam} S.~S., {Allende Prieto}  C., {Anderson} K.~S.~J., {Anderson} S.~F., {Annis} J., {Bahcall} N.~A., {et al.}, 2008, \apjs,  175, 297

\bibitem[{Bacon}, {Massey}, {Refregier}, \&  {Ellis} 2003]{2003MNRAS.344..673B}
{Bacon} D.~J., {Massey} R.~J., {Refregier} A.~R., {Ellis} R.~S., 2003, \mnras,  344, 673

\bibitem[{Ben{\'{\i}}tez} 2000]{2000ApJ...536..571B}
{Ben{\'{\i}}tez} N., 2000, \apj, 536, 571

\bibitem[{Bertin} 2006]{2006ASPC..351..112B}
{Bertin} E., 2006, in Astronomical Society of the Pacific Conference Series,  Vol. 351, Astronomical Data Analysis Software and Systems XV, {Gabriel} C.,  {Arviset} C., {Ponz} D., {Enrique} S., eds., pp. 112--+

\bibitem[{Bertin} \& {Arnouts} 1996]{1996A&AS..117..393B}
{Bertin} E., {Arnouts} S., 1996, \aaps, 117, 393

\bibitem[{Bolzonella}, {Miralles}, \&  {Pell{\'o}} 2000]{2000A&A...363..476B}
{Bolzonella} M., {Miralles} J.-M., {Pell{\'o}} R., 2000, \aap, 363, 476

\bibitem[{Brammer}, {Whitaker}, {van Dokkum},  {Marchesini}, {Labb{\'e}}, {Franx}, {Kriek}, {Quadri}, {Illingworth}, {Lee},  {Muzzin}, \& {Rudnick} 2009]{2009ApJ...706L.173B}
{Brammer} G.~B., {Whitaker} K.~E., {van Dokkum} P.~G., {Marchesini} D.,  {Labb{\'e}} I., {Franx} M., {Kriek} M., {Quadri} R.~F., {et al.}, 2009, \apjl, 706, L173

\bibitem[{Broadhurst}, {Ben{\'{\i}}tez}, {Coe},  {Sharon}, {Zekser}, {White}, {Ford}, {Bouwens}, {Blakeslee}, {Clampin},  {Cross}, {Franx}, {Frye}, {Hartig}, {Illingworth}, {Infante}, {Menanteau},  {Meurer}, {Postman}, {Ardila}, {Bartko}, {Brown}, {Burrows}, {Cheng},  {Feldman}, {Golimowski}, {Goto}, {Gronwall}, {Herranz}, {Holden}, {Homeier},  {Krist}, {Lesser}, {Martel}, {Miley}, {Rosati}, {Sirianni}, {Sparks},  {Steindling}, {Tran}, {Tsvetanov}, \& {Zheng} 2005]{2005ApJ...621...53B}
{Broadhurst} T., {Ben{\'{\i}}tez} N., {Coe} D., {Sharon} K., {Zekser} K.,  {White} R., {Ford} H., {Bouwens} R., {et al.}, 2005, \apj, 621, 53

\bibitem[{Broadhurst}, {Taylor}, \&  {Peacock} 1995]{1995ApJ...438...49B}
{Broadhurst} T.~J., {Taylor} A.~N., {Peacock} J.~A., 1995, \apj, 438, 49

\bibitem[{Brown}, {Ade}, {Bock}, {Bowden}, {Cahill},  {Castro}, {Church}, {Culverhouse}, {Friedman}, {Ganga}, {Gear}, {Gupta},  {Hinderks}, {Kovac}, {Lange}, {Leitch}, {Melhuish}, {Memari}, {Murphy},  {Orlando}, {O'Sullivan}, {Piccirillo}, {Pryke}, {Rajguru}, {Rusholme},  {Schwarz}, {Taylor}, {Thompson}, {Turner}, {Wu}, {Zemcov}, \& {The QUa D  collaboration} 2009]{2009ApJ...705..978B}
{Brown} M.~L., {Ade} P., {Bock} J., {Bowden} M., {Cahill} G., {Castro} P.~G.,  {Church} S., {Culverhouse} T., {et al.}, 2009, \apj, 705, 978

\bibitem[{Brown}, {Taylor}, {Bacon}, {Gray}, {Dye},  {Meisenheimer}, \& {Wolf} 2003]{2003MNRAS.341..100B}
{Brown} M.~L., {Taylor} A.~N., {Bacon} D.~J., {Gray} M.~E., {Dye} S.,  {Meisenheimer} K., {Wolf} C., 2003, \mnras, 341, 100

\bibitem[{Capak}, {Aussel}, {Ajiki}, {McCracken},  {Mobasher}, {Scoville}, {Shopbell}, {Taniguchi}, {Thompson}, {Tribiano},  {Sasaki}, {Blain}, {Brusa}, {Carilli}, {Comastri}, {Carollo}, {Cassata},  {Colbert}, {Ellis}, {Elvis}, {Giavalisco}, {Green}, {Guzzo}, {Hasinger},  {Ilbert}, {Impey}, {Jahnke}, {Kartaltepe}, {Kneib}, {Koda}, {Koekemoer},  {Komiyama}, {Leauthaud}, {Lefevre}, {Lilly}, {Liu}, {Massey}, {Miyazaki},  {Murayama}, {Nagao}, {Peacock}, {Pickles}, {Porciani}, {Renzini}, {Rhodes},  {Rich}, {Salvato}, {Sanders}, {Scarlata}, {Schiminovich}, {Schinnerer},  {Scodeggio}, {Sheth}, {Shioya}, {Tasca}, {Taylor}, {Yan}, \&  {Zamorani} 2007]{2007ApJS..172...99C}
{Capak} P., {Aussel} H., {Ajiki} M., {McCracken} H.~J., {Mobasher} B.,  {Scoville} N., {Shopbell} P., {Taniguchi} Y., {et al.}, 2007, \apjs, 172, 99

\bibitem[{Coe}, {Ben{\'{\i}}tez}, {S{\'a}nchez}, {Jee},  {Bouwens}, \& {Ford} 2006]{2006AJ....132..926C}
{Coe} D., {Ben{\'{\i}}tez} N., {S{\'a}nchez} S.~F., {Jee} M., {Bouwens} R.,  {Ford} H., 2006, \aj, 132, 926

\bibitem[{Fioc} \& {Rocca-Volmerange} 1997]{1997A&A...326..950F}
{Fioc} M., {Rocca-Volmerange} B., 1997, \aap, 326, 950

\bibitem[{Gilmore} \& {Natarajan} 2009]{2009MNRAS.396..354G}
{Gilmore} J., {Natarajan} P., 2009, \mnras, 396, 354

\bibitem[{Grazian}, {Fontana}, {de Santis}, {Nonino},  {Salimbeni}, {Giallongo}, {Cristiani}, {Gallozzi}, \&  {Vanzella} 2006]{2006A&A...449..951G}
{Grazian} A., {Fontana} A., {de Santis} C., {Nonino} M., {Salimbeni} S.,  {Giallongo} E., {Cristiani} S., {Gallozzi} S., {et al.}, 2006, \aap,  449, 951

\bibitem[{Heymans}, {Van Waerbeke}, {Bacon}, {Berge},  {Bernstein}, {Bertin}, {Bridle}, {Brown}, {Clowe}, {Dahle}, {Erben}, {Gray},  {Hetterscheidt}, {Hoekstra}, {Hudelot}, {Jarvis}, {Kuijken}, {Margoniner},  {Massey}, {Mellier}, {Nakajima}, {Refregier}, {Rhodes}, {Schrabback}, \&  {Wittman} 2006]{2006MNRAS.368.1323H}
{Heymans} C., {Van Waerbeke} L., {Bacon} D., {Berge} J., {Bernstein} G.,  {Bertin} E., {Bridle} S., {Brown} M.~L., {et al.}, 2006, \mnras,  368, 1323

\bibitem[{Ilbert}, {Capak}, {Salvato}, {Aussel},  {McCracken}, {Sanders}, {Scoville}, {Kartaltepe}, {Arnouts}, {Floc'h},  {Mobasher}, {Taniguchi}, {Lamareille}, {Leauthaud}, {Sasaki}, {Thompson},  {Zamojski}, {Zamorani}, {Bardelli}, {Bolzonella}, {Bongiorno}, {Brusa},  {Caputi}, {Carollo}, {Contini}, {Cook}, {Coppa}, {Cucciati}, {de la Torre},  {de Ravel}, {Franzetti}, {Garilli}, {Hasinger}, {Iovino}, {Kampczyk},  {Kneib}, {Knobel}, {Kovac}, {LeBorgne}, {LeBrun}, {F{\`e}vre}, {Lilly},  {Looper}, {Maier}, {Mainieri}, {Mellier}, {Mignoli}, {Murayama}, {Pell{\`o}},  {Peng}, {P{\'e}rez-Montero}, {Renzini}, {Ricciardelli}, {Schiminovich},  {Scodeggio}, {Shioya}, {Silverman}, {Surace}, {Tanaka}, {Tasca}, {Tresse},  {Vergani}, \& {Zucca} 2009]{2009ApJ...690.1236I}
{Ilbert} O., {Capak} P., {Salvato} M., {Aussel} H., {McCracken} H.~J.,  {Sanders} D.~B., {Scoville} N., {Kartaltepe} J., {et al.}, 2009, \apj,  690, 1236

\bibitem[{Inada}, {Oguri}, {Pindor}, {Hennawi}, {Chiu},  {Zheng}, {Ichikawa}, {Gregg}, {Becker}, {Suto}, {Strauss}, {Turner},  {Keeton}, {Annis}, {Castander}, {Eisenstein}, {Frieman}, {Fukugita}, {Gunn},  {Johnston}, {Kent}, {Nichol}, {Richards}, {Rix}, {Sheldon}, {Bahcall},  {Brinkmann}, {Ivezi{\'c}}, {Lamb}, {McKay}, {Schneider}, \&  {York} 2003]{2003Natur.426..810I}
{Inada} N., {Oguri} M., {Pindor} B., {Hennawi} J.~F., {Chiu} K., {Zheng} W.,  {Ichikawa} S.-I., {Gregg} M.~D., {et al.}, 2003, \nat, 426, 810

\bibitem[{Jain}, {Connolly}, \&  {Takada} 2007]{2007JCAP...03..013J}
{Jain} B., {Connolly} A., {Takada} M., 2007, \jcap, 3, 13

\bibitem[{Jain} \& {Taylor} 2003]{2003PhRvL..91n1302J}
{Jain} B., {Taylor} A., 2003, Physical Review Letters, 91, 141302

\bibitem[{Kaiser}, {Squires}, \&  {Broadhurst} 1995]{1995ApJ...449..460K}
{Kaiser} N., {Squires} G., {Broadhurst} T., 1995, \apj, 449, 460

\bibitem[{Kitching}, {Heavens}, {Taylor}, {Brown},  {Meisenheimer}, {Wolf}, {Gray}, \& {Bacon} 2007]{2007MNRAS.376..771K}
{Kitching} T.~D., {Heavens} A.~F., {Taylor} A.~N., {Brown} M.~L.,  {Meisenheimer} K., {Wolf} C., {Gray} M.~E., {Bacon} D.~J., 2007, \mnras, 376,  771

\bibitem[{Kotulla}, {Fritze}, {Weilbacher}, \&  {Anders} 2009]{2009MNRAS.396..462K}
{Kotulla} R., {Fritze} U., {Weilbacher} P., {Anders} P., 2009, \mnras, 396, 462

\bibitem[{Massey}, {Rhodes}, {Leauthaud}, {Capak},  {Ellis}, {Koekemoer}, {R{\'e}fr{\'e}gier}, {Scoville}, {Taylor}, {Albert},  {Berg{\'e}}, {Heymans}, {Johnston}, {Kneib}, {Mellier}, {Mobasher},  {Semboloni}, {Shopbell}, {Tasca}, \& {Van Waerbeke} 2007]{2007ApJS..172..239M}
{Massey} R., {Rhodes} J., {Leauthaud} A., {Capak} P., {Ellis} R., {Koekemoer}  A., {R{\'e}fr{\'e}gier} A., {Scoville} N., {et al.}, 2007, \apjs, 172, 239

\bibitem[{Medezinski}, {Broadhurst}, {Umetsu},  {Coe}, {Ben{\'{\i}}tez}, {Ford}, {Rephaeli}, {Arimoto}, \&  {Kong} 2007]{2007ApJ...663..717M}
{Medezinski} E., {Broadhurst} T., {Umetsu} K., {Coe} D., {Ben{\'{\i}}tez} N.,  {Ford} H., {Rephaeli} Y., {Arimoto} N., {et al.}, 2007, \apj, 663, 717

\bibitem[{Medezinski}, {Broadhurst}, {Umetsu},  {Oguri}, {Rephaeli}, \& {Ben{\'{\i}}tez} 2010]{2010MNRAS.405..257M}
{Medezinski} E., {Broadhurst} T., {Umetsu} K., {Oguri} M., {Rephaeli} Y.,  {Ben{\'{\i}}tez} N., 2010, \mnras, 405, 257

\bibitem[{Miyazaki}, {Komiyama}, {Sekiguchi},  {Okamura}, {Doi}, {Furusawa}, {Hamabe}, {Imi}, {Kimura}, {Nakata}, {Okada},  {Ouchi}, {Shimasaku}, {Yagi}, \& {Yasuda} 2002]{2002PASJ...54..833M}
{Miyazaki} S., {Komiyama} Y., {Sekiguchi} M., {Okamura} S., {Doi} M.,  {Furusawa} H., {Hamabe} M., {Imi} K., {et al.}, 2002, \pasj, 54, 833

\bibitem[{Oguri}, {Inada}, {Keeton}, {Pindor},  {Hennawi}, {Gregg}, {Becker}, {Chiu}, {Zheng}, {Ichikawa}, {Suto}, {Turner},  {Annis}, {Bahcall}, {Brinkmann}, {Castander}, {Eisenstein}, {Frieman},  {Goto}, {Gunn}, {Johnston}, {Kent}, {Nichol}, {Richards}, {Rix}, {Schneider},  {Sheldon}, \& {Szalay} 2004]{2004ApJ...605...78O}
{Oguri} M., {Inada} N., {Keeton} C.~R., {Pindor} B., {Hennawi} J.~F., {Gregg}  M.~D., {Becker} R.~H., {Chiu} K., {et al.}, 2004, \apj, 605,  78

\bibitem[{Perlmutter}, {Aldering}, {Goldhaber},  {Knop}, {Nugent}, {Castro}, {Deustua}, {Fabbro}, {Goobar}, {Groom}, {Hook},  {Kim}, {Kim}, {Lee}, {Nunes}, {Pain}, {Pennypacker}, {Quimby}, {Lidman},  {Ellis}, {Irwin}, {McMahon}, {Ruiz-Lapuente}, {Walton}, {Schaefer}, {Boyle},  {Filippenko}, {Matheson}, {Fruchter}, {Panagia}, {Newberg}, {Couch}, \& {The  Supernova Cosmology Project} 1999]{1999ApJ...517..565P}
{Perlmutter} S., {Aldering} G., {Goldhaber} G., {Knop} R.~A., {Nugent} P.,  {Castro} P.~G., {Deustua} S., {Fabbro} S., {et al.}, 1999,  \apj, 517, 565

\bibitem[{Riess}, {Filippenko}, {Challis},  {Clocchiatti}, {Diercks}, {Garnavich}, {Gilliland}, {Hogan}, {Jha},  {Kirshner}, {Leibundgut}, {Phillips}, {Reiss}, {Schmidt}, {Schommer},  {Smith}, {Spyromilio}, {Stubbs}, {Suntzeff}, \&  {Tonry} 1998]{1998AJ....116.1009R}
{Riess} A.~G., {Filippenko} A.~V., {Challis} P., {Clocchiatti} A., {Diercks}  A., {Garnavich} P.~M., {Gilliland} R.~L., {Hogan} C.~J., {et al.}, 1998, \aj, 116, 1009

\bibitem[{Santini}, {Fontana}, {Grazian}, {Salimbeni},  {Fiore}, {Fontanot}, {Boutsia}, {Castellano}, {Cristiani}, {de Santis},  {Gallozzi}, {Giallongo}, {Menci}, {Nonino}, {Paris}, {Pentericci}, \&  {Vanzella} 2009]{2009A&A...504..751S}
{Santini} P., {Fontana} A., {Grazian} A., {Salimbeni} S., {Fiore} F.,  {Fontanot} F., {Boutsia} K., {Castellano} M., {et al.}, 2009, \aap, 504, 751

\bibitem[{Schrabback}, {Hartlap}, {Joachimi},  {Kilbinger}, {Simon}, {Benabed}, {Brada{\v c}}, {Eifler}, {Erben},  {Fassnacht}, {High}, {Hilbert}, {Hildebrandt}, {Hoekstra}, {Kuijken},  {Marshall}, {Mellier}, {Morganson}, {Schneider}, {Semboloni}, {van Waerbeke},  \& {Velander} 2010]{2010A&A...516A..63S}
{Schrabback} T., {Hartlap} J., {Joachimi} B., {Kilbinger} M., {Simon} P.,  {Benabed} K., {Brada{\v c}} M., {Eifler} T., {et al.}, 2010, \aap, 516, A63+

\bibitem[{Seitz} \& {Schneider} 1997]{1997A&A...318..687S}
{Seitz} C., {Schneider} P., 1997, \aap, 318, 687

\bibitem[{Sharon}, {Ofek}, {Smith}, {Broadhurst},  {Maoz}, {Kochanek}, {Oguri}, {Suto}, {Inada}, \&  {Falco} 2005]{2005ApJ...629L..73S}
{Sharon} K., {Ofek} E.~O., {Smith} G.~P., {Broadhurst} T., {Maoz} D.,  {Kochanek} C.~S., {Oguri} M., {Suto} Y., {et al.}, 2005,  \apjl, 629, L73

\bibitem[{Soucail}, {Kneib}, \&  {Golse} 2004]{2004A&A...417L..33S}
{Soucail} G., {Kneib} J., {Golse} G., 2004, \aap, 417, L33

\bibitem[{Spergel}, {Bean}, {Dor{\'e}}, {Nolta},  {Bennett}, {Dunkley}, {Hinshaw}, {Jarosik}, {Komatsu}, {Page}, {Peiris},  {Verde}, {Halpern}, {Hill}, {Kogut}, {Limon}, {Meyer}, {Odegard}, {Tucker},  {Weiland}, {Wollack}, \& {Wright} 2007]{2007ApJS..170..377S}
{Spergel} D.~N., {Bean} R., {Dor{\'e}} O., {Nolta} M.~R., {Bennett} C.~L.,  {Dunkley} J., {Hinshaw} G., {Jarosik} N., {et al.}, 2007, \apjs, 170, 377

\bibitem[{Steidel}, {Adelberger}, {Giavalisco},  {Dickinson}, \& {Pettini} 1999]{1999ApJ...519....1S}
{Steidel} C.~C., {Adelberger} K.~L., {Giavalisco} M., {Dickinson} M., {Pettini}  M., 1999, \apj, 519, 1

\bibitem[{Taylor}, {Bacon}, {Gray}, {Wolf},  {Meisenheimer}, {Dye}, {Borch}, {Kleinheinrich}, {Kovacs}, \&  {Wisotzki} 2004]{2004MNRAS.353.1176T}
{Taylor} A.~N., {Bacon} D.~J., {Gray} M.~E., {Wolf} C., {Meisenheimer} K.,  {Dye} S., {Borch} A., {Kleinheinrich} M., {et al.}, 2004,  \mnras, 353, 1176

\bibitem[{Taylor}, {Kitching}, {Bacon}, \&  {Heavens} 2007]{2007MNRAS.374.1377T}
{Taylor} A.~N., {Kitching} T.~D., {Bacon} D.~J., {Heavens} A.~F., 2007, \mnras,  374, 1377

\bibitem[{Umetsu}, {Birkinshaw}, {Liu}, {Wu},  {Medezinski}, {Broadhurst}, {Lemze}, {Zitrin}, {Ho}, {Huang}, {Koch}, {Liao},  {Lin}, {Molnar}, {Nishioka}, {Wang}, {Altamirano}, {Chang}, {Chang}, {Chang},  {Chen}, {Han}, {Huang}, {Hwang}, {Jiang}, {Kesteven}, {Kubo}, {Li},  {Martin-Cocher}, {Oshiro}, {Raffin}, {Wei}, \&  {Wilson} 2009]{2009ApJ...694.1643U}
{Umetsu} K., {Birkinshaw} M., {Liu} G.-C., {Wu} J.-H.~P., {Medezinski} E.,  {Broadhurst} T., {Lemze} D., {Zitrin} A., {et al.}, 2009, \apj, 694, 1643

\bibitem[{Umetsu} \& {Broadhurst} 2008]{2008ApJ...684..177U}
{Umetsu} K., {Broadhurst} T., 2008, \apj, 684, 177

\bibitem[{Umetsu}, {Medezinski}, {Broadhurst},  {Zitrin}, {Okabe}, {Hsieh}, \& {Molnar} 2010]{2010ApJ...714.1470U}
{Umetsu} K., {Medezinski} E., {Broadhurst} T., {Zitrin} A., {Okabe} N., {Hsieh}  B., {Molnar} S.~M., 2010, \apj, 714, 1470

\bibitem[{Wittman}, {Tyson}, {Margoniner}, {Cohen}, \&  {Dell'Antonio} 2001]{2001ApJ...557L..89W}
{Wittman} D., {Tyson} J.~A., {Margoniner} V.~E., {Cohen} J.~G., {Dell'Antonio}  I.~P., 2001, \apjl, 557, L89

\bibitem[{Yagi}, {Kashikawa}, {Sekiguchi}, {Doi},  {Yasuda}, {Shimasaku}, \& {Okamura} 2002]{2002AJ....123...66Y}
{Yagi} M., {Kashikawa} N., {Sekiguchi} M., {Doi} M., {Yasuda} N., {Shimasaku}  K., {Okamura} S., 2002, \aj, 123, 66

\bibitem[{Zacharias}, {Monet}, {Levine}, {Urban},  {Gaume}, \& {Wycoff} 2004]{2004AAS...205.4815Z}
{Zacharias} N., {Monet} D.~G., {Levine} S.~E., {Urban} S.~E., {Gaume} R.,  {Wycoff} G.~L., 2004, in Bulletin of the American Astronomical Society,  Vol.~36, Bulletin of the American Astronomical Society, pp. 1418--+

\bibitem[{Zitrin} \& {Broadhurst} 2009]{2009ApJ...703L.132Z}
{Zitrin} A., {Broadhurst} T., 2009, \apjl, 703, L132

\bibitem[{Zitrin}, {Broadhurst},  {Rephaeli}, \& {Sadeh} 2009a]{2009ApJ...707L.102Z}
{Zitrin} A., {Broadhurst} T., {Rephaeli} Y., {Sadeh} S., 2009a,  \apjl, 707, L102

\bibitem[{Zitrin}, {Broadhurst},  {Umetsu}, {Coe}, {Ben{\'{\i}}tez}, {Ascaso}, {Bradley}, {Ford}, {Jee},  {Medezinski}, {Rephaeli}, \& {Zheng} 2009b]{2009MNRAS.396.1985Z}
{Zitrin} A., {Broadhurst} T., {Umetsu} K., {Coe} D., {Ben{\'{\i}}tez} N.,  {Ascaso} B., {Bradley} L., {Ford} H., {et al.}, 2009b, \mnras, 396, 1985

\end{thebibliography}

\end{document}